\documentclass[journal,onecolumn]{IEEEtran}
\usepackage[ruled,linesnumbered]{algorithm2e}

\renewcommand{\IEEEQED}{\IEEEQEDclosed}
\usepackage{color}

\usepackage{subfigure}
\usepackage{graphicx,cite,epsfig,amssymb,amsmath,multirow,lettrine,flushend,extarrows}

\usepackage{threeparttable}

\begin{document}



\title{Multi-Agent Deep Reinforcement Learning for HVAC Control in Commercial Buildings}

\author{{Liang~Yu,~\IEEEmembership{Member,~IEEE}, Yi~Sun, Zhanbo Xu,~\IEEEmembership{Member,~IEEE}, Chao Shen,~\IEEEmembership{Senior Member,~IEEE},\\ Dong Yue,~\IEEEmembership{Senior Member,~IEEE}, Tao~Jiang,~\IEEEmembership{Fellow,~IEEE}, Xiaohong~Guan,~\IEEEmembership{Fellow,~IEEE}}
\thanks{\newline L. Yu is with the College of Automation \& College of Artificial Intelligence, Nanjing University of Posts and Telecommunications, Nanjing 210003, China, and also with Systems Engineering Institute, Ministry of Education Key Lab for Intelligent Networks and Network Security, Xi'an Jiaotong University, Xi'an 710049, China. (email: liang.yu@njupt.edu.cn) \newline
Y. Sun is with the college of Internet of Things, Nanjing University of Posts and Telecommunications, Nanjing 210003, China. (email: yi\_sun99@yeah.net) \newline
Z. Xu, C. Shen, and X. Guan are with Systems Engineering Institute, Ministry of Education Key Lab for Intelligent Networks and Network Security, Xi'an Jiaotong University, Xi'an 710049, China. (email: \{zbxu,cshen,xhguan\}@sei.xjtu.edu.cn) \newline
D. Yue is with the Institute of Advanced Technology, Nanjing University of Posts and Telecommunications, Nanjing 210003, China. (email: medongy@vip.163.com) \newline
T. Jiang is with Wuhan National Laboratory for Optoelectronics, School of Electronic Information and Communications, Huazhong University of Science and Technology, Wuhan 430074, China. (email: Tao.Jiang@ieee.org) \newline
}}

\markboth{IEEE Transactions on Smart Grid,~Vol.~XX, No.~XX, Month~2020}%
{Liang \MakeLowercase{\textit{et al.}}: Multi-Agent Deep Reinforcement Learning for HVAC Control in Commercial Buildings}

\maketitle

\begin{abstract}
In commercial buildings, about 40\%-50\% of the total electricity consumption is attributed to Heating, Ventilation, and Air Conditioning (HVAC) systems, which places an economic burden on building operators. In this paper, we intend to minimize the energy cost of an HVAC system in a multi-zone commercial building with the consideration of random zone occupancy, thermal comfort, and indoor air quality comfort. Due to the existence of unknown thermal dynamics models, parameter uncertainties (e.g., outdoor temperature, electricity price, and number of occupants), spatially and temporally coupled constraints associated with indoor temperature and $\text{CO}_2$ concentration, a large discrete solution space, and a non-convex and non-separable objective function, it is very challenging to achieve the above aim. To this end, the above energy cost minimization problem is reformulated as a Markov game. Then, an HVAC control algorithm is proposed to solve the Markov game based on multi-agent deep reinforcement learning with attention mechanism. The proposed algorithm does not require any prior knowledge of uncertain parameters and can operate without knowing building thermal dynamics models. Simulation results based on real-world traces show the effectiveness, robustness and scalability of the proposed algorithm.
\end{abstract}


\section*{Nomenclature}
\textbf{Indices}
\begin{description}
\item[$t$]~Time slot index.
\item[$i,z$]~Zone index.
\end{description}

\textbf{Parameters and Constants}
\begin{description}
\item[$N$]~Total number of zones.
\item[$L$]~Total number of time slots.
\item[$T_i^{\min}$]~Minimum acceptable indoor temperature ($^oC$).
\item[$T_i^{\max}$]~Maximum acceptable indoor temperature ($^oC$).
\item[$\mathcal{N}_i$]~The set of neighbors of zone $i$.
\item[$O_{i}^{\max}$]~Acceptable $\text{CO}_2$ concentration in zone $i$ ($\text{ppm}$).
\item[$\tau$]~Time slot length (min).
\item[$\kappa$]~The air density ($g/m^3$).
\item[$\vartheta_i$]~The volume of zone $i$ ($m^3$).
\item[$\chi$]~The $\text{CO}_2$ generation rate per person ($L/s$).
\item[$M$]~Number of discrete levels related to $m_{i,t}$.
\item[$Z$]~Number of discrete levels related to $\sigma_{t}$.
\item[$\mu$]~Fan power consumption coefficient ($\text{Watt}/(g/s)^3$).
\item[$C_a$]~The specific heat of the air ($J/g/^oC$).
\item[$\eta$]~The efficiency factor of the cooling coil.
\item[$COP$]~Coefficient of performance related to the chiller.
\item[$T_s$]~Supply air temperature of the VFD fan ($^oC$).
\item[$\theta$]~The weight parameter of actor network.
\item[$\psi$]~The weight parameter of critic network.
\end{description}

\textbf{Variables}
\begin{description}
\item[$T_{i,t}$]~Indoor temperature of zone $i$ at slot $t$ ($^oC$).
\item[$K_{i,t}$]~The number of occupants in zone $i$ at slot $t$.
\item[$T_t^{\text{out}}$]~Outdoor temperature at slot $t$ ($^oC$).
\item[$m_{i,t}$]~Air supply rate of zone $i$ at slot $t$ ($g/s$).
\item[$\varsigma_{i,t}$]~Thermal disturbance in zone $i$ at slot $t$ ($\text{Watts}$).
\item[$O_{i,t}$]~$\text{CO}_2$ concentration in zone $i$ at slot $t$ ($\text{ppm}$).
\item[$O_{t}^{\text{mix}}$]~$\text{CO}_2$ concentration of the mixed air at slot $t$ ($\text{ppm}$).
\item[$O_{t}^{\text{out}}$]~Outside $\text{CO}_2$ concentration at slot $t$ ($\text{ppm}$).
\item[$\sigma_t$]~The damper position in the AHU.
\item[$\lambda_t$]~Electricity price ($\text{RMB}/kWh$).
\item[$T_t^{\text{mix}}$]~Mixed air temperature ($^oC$).
\item[$p_t$]~Power consumption of the cooling coil ($\text{Watts}$).
\item[$\Phi_{1,t}$]~The energy cost related to the supply fan ($\text{RMB}$).
\item[$\Phi_{2,t}$]~The energy cost related to the cooling coil ($\text{RMB}$).
\item[$I_{i,t}$]~Occupancy state indicator of zone $i$ at slot $t$.
\end{description}

\section{Introduction}\label{s1}
As electricity consumers in smart grid, buildings are responsible for a large portion of the total electricity consumption in a country. For example, residential buildings and commercial buildings accounted for 38.7\% and 35.5\% of the total electricity usage of U.S. in 2010, respectively\cite{Book2011}. In commercial buildings (e.g., offices, stores, restaurants, warehouses, other buildings used for commercial purposes, and government buildings\cite{Book2011}), about 40\%-50\% of the total electricity consumption is attributed to Heating, Ventilation, and Air Conditioning (HVAC) systems, which places an economic burden on building operators. Fortunately, the slow thermal dynamic characteristics of buildings make HVAC systems perfect candidates for demand side management due to building thermal inertia\cite{Liu2017}. Under some operational constraints (e.g., comfortable indoor temperature range and comfortable indoor air quality), HVAC systems can be scheduled flexibly to save energy cost as a response to dynamic prices\cite{LiangTSG2017,Cao2012,Tsui2012}. In other words, the energy consumption of HVAC systems will decrease during high/peak price hours and increase during low price hours, which can also offer many benefits for microgrid (i.e., a low voltage distribution network comprising various distributed generation, storage devices, and responsive loads\cite{Liu2017}). For example, incorporating building HVAC control in microgrid scheduling\cite{Liu2017} and planning\cite{ZhangSPEC2019} are beneficial to reduce operation cost and total annualized cost (including investment cost and operation cost), respectively.

Many approaches have been proposed to minimize HVAC energy cost in commercial buildings under dynamic prices with the consideration of thermal comfort and/or indoor air quality comfort, e.g., model predictive control (MPC)\cite{Mantovani2015}, stochastic MPC\cite{Ma2015}, event-based approach\cite{Wu2016}, distributed MPC\cite{ZhiWang2017}, Lyapunov optimization techniques\cite{Yu2018JIOT}\cite{LiangTSG2019}, convex optimization\cite{Hao2017}, mixed-integer linear programming\cite{Kim2018}, Lagrangian relaxation approach\cite{Xu2017}, and non-linear optimization\cite{Vishwanath2019}. Though some advances have been made in above-mentioned studies, they have three drawbacks. Firstly, they need to know building thermal dynamics models. Since building indoor temperature depends on many factors, it is very challenging to develop a building thermal dynamics model that is accurate and efficient enough for HVAC control\cite{Wei2017}\cite{Yoon2019}. Moreover, the performances or premises of model-based methods depend on specific building environment and their generalities are limited when confronted with various building environments\cite{LiangTSG2019}\cite{Gao2019}. Secondly, they need to predict uncertain parameters or know explicit models (e.g., probability distribution) of representing uncertainties. When prediction errors are large, algorithmic performance will be affected\cite{Mantovani2015}\cite{Ma2015}\cite{Chen2019}. Thirdly, the above-mentioned methods do not support on-line decision-making for large-scale solution space\cite{Mocanu2019}. To be specific, any time when an optimization is needed, these methods have to compute completely or partially all the possible solutions and choose the best one. When the solution space is very large, the computation process is time-consuming.

To overcome the above-mentioned drawbacks, learning-based techniques can be adopted, e.g., reinforcement learning\cite{Sutton2018,Deng2020,Canteli2019,Han2019} and deep reinforcement learning (DRL)\cite{Mnih2015,Wang2020,Wang2019}. To be specific, they can fully exploit the information in interactions with the building environment to learn an optimal policy without knowing building thermal dynamics models. Once the learning process is finished, the obtained policy can be used for end-to-end complex decision-making, which can generate an optimal action instantly (e.g., within several milliseconds) given a system state without knowing any prior knowledge of uncertain parameters. Although reinforcement learning based methods in \cite{LuTSG2019}\cite{Ruelens2017} do not require the prior knowledge of building thermal dynamics models, they are known to be unstable or even to diverge when a nonlinear function approximator (e.g., a deep neural network) is used to represent the action-value function\cite{Mnih2015}. To efficiently handle the problem with large and continuous state space, many DRL-based HVAC control methods have been proposed. For example, Wei \emph{et al.}\cite{Wei2017} proposed an HVAC control method based on Deep Q-Network (DQN) to save energy cost while maintaining the room temperature within the desired range. In \cite{Gao2019}, Gao \emph{et al.} presented an HVAC control method based on Deep Deterministic Policy Gradients (DDPG) to minimize energy consumption and thermal discomfort in a laboratory. In \cite{Zhang2019}, Zhang \emph{et al.} proposed an HVAC control framework based on Asynchronous Advantage Actor Critic (A3C), which jointly optimizes energy demand and thermal comfort in an office building. In \cite{Ding2019}, Ding \emph{et al.} developed a control method based on Branching Dueling Q-network (BDQ) for four building subsystems, including HVAC, lighting, blind and window systems. In \cite{Mocanu2019}, Mocanu \emph{et al.} proposed two optimization methods based on DQN and Deep Policy Gradient (DPG) to minimize energy cost by scheduling a set of electrical devices in residential buildings. In \cite{YuIoT2019}, Yu \emph{et al.} proposed a DDPG-based energy management algorithm to minimize the energy cost of a smart home with energy storage systems and HVAC systems. In \cite{He2020}, Li \emph{et al.} proposed a trust region policy optimization (TRPO) based demand response strategy for optimal scheduling of home appliances. Although the above DRL-based HVAC control methods are effective, they did not jointly consider random zone occupancy, thermal comfort and indoor air quality comfort in a multi-zone commercial building.

Based on the above observation, this paper intends to minimize HVAC energy cost in a multi-zone commercial building under dynamic prices, with the consideration of random zone occupancy, thermal comfort and indoor air quality comfort in the absence of building thermal dynamics models. To be specific, air supply rate in each zone and the damper position in the air handling unit are jointly determined to minimize the long-term HVAC energy cost while maintaining comfortable temperature and $\text{CO}_2$ concentration ranges. However, several challenges are involved in achieving the above aim. Firstly, building thermal dynamics models are unknown. Secondly, there are spatially and temporally constraints associated with indoor temperature and $\text{CO}_2$ concentration. Thirdly, the objective function is non-convex and non-separable. Fourthly, the solution space at each time slot is extremely large. For example, the total number of solutions would be $10^{51}$ if 10 discrete variable levels and 50 zones are considered. Finally, there are some uncertain parameters (e.g., electricity price, outdoor temperature, number of occupants). When taking above-mentioned challenges into consideration, all existing model-based or model-free methods are not applicable to our problem. To this end, we propose an HVAC control algorithm for the multi-zone commercial building based on multi-agent deep reinforcement learning (MADRL) with attention mechanism\cite{Iqbal2019}, which supports flexible and scalable coordination among different agents.

The main contributions of this paper are summarized as follows.

\begin{itemize}
  \item By taking random zone occupancy, comfortable temperature range, and comfortable $\text{CO}_2$ concentration range into consideration, we first formulate a long-term HVAC energy cost minimization problem related to a multi-zone commercial building without knowing building thermal dynamics models. Then, we reformulate the optimization problem as a Markov game, where environment state, action, and reward function are designed.
  \item We propose a scalable HVAC control algorithm to solve the Markov game based on multi-agent deep reinforcement learning with attention mechanism. Note that the proposed algorithm is model-free and does not require any prior knowledge of uncertain parameters.
  \item Simulation results based on real-world traces show the effectiveness, robustness, and scalability of the proposed algorithm. When 30 zones are considered, the proposed algorithm can reduce average energy cost by 56.50\%-75.25\% compared with other baselines while still maintaining the comfort for occupants.
\end{itemize}

The rest of this paper is organized as follows. In Section~\ref{s2}, we describe the system model and provide problem formulation. In Section~\ref{s3}, we propose a MADRL-based HVAC control algorithm. In Section~\ref{s4}, we conduct performance evaluation. Finally, conclusions are made in Section~\ref{s5}.

\section{System Model and Problem Formulation}\label{s2}
\begin{figure}[!htb]
\centering
\includegraphics[scale=0.58]{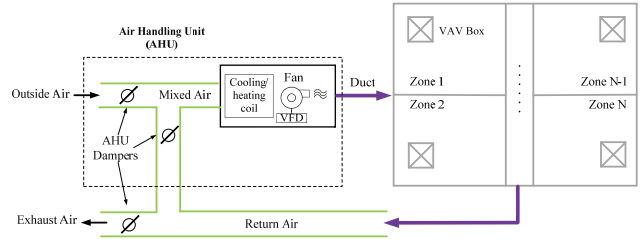}
\caption{Typical commercial HVAC system.}\label{fig_1}
\end{figure}

We consider a commercial building with $N$ thermal zones and indoor temperatures in these zones could be adjusted by an HVAC system as shown in Fig.~\ref{fig_1}. A typical commercial HVAC system consists of an air handling unit (AHU) for the whole building and a set of variable air volume (VAV) boxes for each zone\cite{Hao2017}. The AHU is composed of dampers, a cooling/heating coil, and a variable frequency drive (VFD) fan. The dampers could mix the outside fresh air with the air returned from each zone to satisfy the ventilation requirement of each zone. The cooling/heating coil cools down/heats up the mixed air, while the VFD fan could deliver the mixed air to the VAV box in each zone. Without loss of generality, we focus on the case that all zones need cooling. In addition, we assume that the HVAC system operates in slotted time (i.e., $t=\{1,\cdots,L\}$). In the following parts, we first describe the HVAC control and cost models. Then, we formulate an energy cost minimization problem and its variant.

\subsection{HVAC Control Model}\label{s2_1}
In this paper, we intend to adjust the damper position in the AHU and air supply rate in each zone so that thermal comfort and indoor air quality comfort could be maintained. Similar to \cite{Wei2017}, a comfortable temperature range is adopted as the representation of thermal comfort for simplicity. Let $T_{i,t}$ be the indoor temperature of zone $i$ at slot $t$. Then, we have
\begin{align} \label{f_1}
T_i^{\min}\leq T_{i,t}\leq T_i^{\max},~\forall~i,t,~K_{i,t}>0,
\end{align}
where $K_{i,t}$ denotes the number of occupants in zone $i$ at slot $t$, while $T_i^{\min}$ and $T_i^{\max}$ denote the minimum and maximum acceptable indoor temperature at zone $i$ respectively.

According to \cite{Hao2017}\cite{Zhang2017}, indoor temperature in zone $i$ at slot $t+1$ (i.e., $T_{i,t+1}$) depends on many factors, e.g., the indoor temperature in zone $i$ as well as its adjacent zones at slot $t$ (i.e., $T_{z,t}$), outdoor temperature at slot $t$ (i.e., $T_{t}^{\text{out}}$), air supply rate in zone $i$ at slot $t$ (i.e., $m_{i,t}$), and thermal disturbance in zone $i$ at slot $t$ (i.e., $\varsigma_{i,t}$). Then, we have
\begin{align} \label{f_2}
T_{i,t+1}=\mathcal{F}(T_{i,t},~T_{z,t}|_{\forall z\in \mathcal{N}_i},~T_{t}^{\text{out}},~m_{i,t},\varsigma_{i,t}),
\end{align}
where $\mathcal{N}_i$ is the set of neighbors of zone $i$. Since it is difficult to develop a building thermal dynamics model that is both accurate and efficient enough for effective HVAC control\cite{Wei2017}, we assume that the explicit model of $\mathcal{F}(\cdot)$ is unknown.

As for indoor air quality comfort, $\text{CO}_2$ concentration level is selected as its representation\cite{Wang2012TSG}. Since high $\text{CO}_2$ concentration level in a zone is harmful to occupants' health and their productivity, it should be controlled below a threshold value $O_i^{\max}$\cite{Gaun2010TSG}. Then, we have
\begin{align} \label{f_3}
O_{i,t} \leq O_i^{\max},~\forall~i,t,~K_{i,t}>0,
\end{align}
where $O_{i,t}$ denotes the $\text{CO}_2$ concentration level of zone $i$ at slot $t$.

According to \cite{LiangTSG2019}, the dynamics of $O_{i,t}$ can be described by
\begin{align} \label{f_4}
O_{i,t+1}=(1-\frac{m_{i,t}\tau}{\kappa\vartheta_i})O_{i,t}+\frac{m_{i,t}\tau}{\kappa\vartheta_i}O_t^{\text{mix}}+\frac{K_{i,t}\tau\chi}{\vartheta_i},
\end{align}
where $\tau$ denotes the duration of a time slot, $\kappa$ is the air density, $\vartheta_i$ is the volume of zone $i$, $O_t^{\text{mix}}$ is the $\text{CO}_2$ concentration of the mixed air at slot $t$, and $\chi$ is the $\text{CO}_2$ generation rate per person.

In \eqref{f_4}, $O_t^{\text{mix}}$ can be described by $(1-\sigma_t)O_t^{\text{out}}+\sigma_t\frac{\sum\nolimits_{i}O_{i,t}m_{i,t}}{\sum\nolimits_{i}m_{i,t}}$\cite{LiangTSG2019}, where $m_{i,t}$ is determined by the damper position in the VAV terminal box. Typically, its value can be selected from $M$ discrete levels\cite{Wei2017}. Then, we have
\begin{align} \label{f_5}
m_{i,t} \in \{m_{i}^1,~m_{i}^2,\cdots,~m_{i}^{M}\},~\forall~i,~t.
\end{align}

Similarly, the damper position in the AHU $\sigma_t$ can be chosen from $Z$ discrete values, i.e.,
\begin{align} \label{f_6}
\sigma_t \in \{\sigma^1,~\sigma^2,\cdots,~\sigma^Z\},~\forall~t.
\end{align}

\subsection{HVAC Energy Cost Model}\label{s2_2}
According to \cite{LiangTSG2019}\cite{Hao2017}, HVAC energy cost consists of two parts, which are related to the supply fan and the cooling coil. Moreover, power consumption associated with the supply fan could be approximated by $\mu(\sum\nolimits_i m_{i,t})^3$\cite{Zhang2017}, where $\mu$ is a coefficient. Then, the energy cost related to the supply fan is given as follows,
\begin{align} \label{f_7}
\Phi_{1,t}=\mu(\sum\nolimits_{i=1}^N m_{i,t})^3\lambda_t\tau,~\forall~t,
\end{align}
where $\lambda_t$ denotes electricity price at slot $t$.

Similarly, the power consumption of the cooling coil $p_t$ could be described by\cite{Hao2017}
\begin{align} \label{f_8}
p_{t}=\frac{C_a\sum\nolimits_i m_{i,t}(T_t^{\text{mix}}-T_s)}{\eta COP},~\forall~t,
\end{align}
where $C_a$ denotes the specific heat of the air, $\eta$ is the efficiency factor of the cooling coil, $COP$ is the coefficient of performance related to the chiller, $T_t^{\text{mix}}=\sigma_t\frac{\sum\nolimits_i m_{i,t}T_{i,t}}{\sum\nolimits_i m_{i,t}}+(1-\sigma_t)T_t^{\text{out}}$ is the mixed air temperature, $T_s$ is the supply air temperature of the VFD fan.

Substituting $T_t^{\text{mix}}$ into \eqref{f_8}, $p_{t}$ could be rewritten as follows,
\begin{align} \label{f_9}
p_{t}=\sum\nolimits_i p_{i,t},
\end{align}
where $p_{i,t}=m_{i,t}\frac{C_a}{\eta COP}(\sigma_t T_{i,t}+(1-\sigma_t)T_t^{\text{out}}-T_s)$. Then, the energy cost related to the cooling coil is given by
\begin{align} \label{f_10}
\Phi_{2,t}=p_{t}\lambda_t\tau,~\forall~t.
\end{align}

\subsection{Energy Cost Minimization Problem}\label{s2_3}
Based on above-mentioned models, we can formulate a stochastic program that minimizes the long-term HVAC energy cost as follows,
\begin{subequations}\label{f_11}
\begin{align}
(\textbf{P1})~&\min_{m_{i,t},\sigma_t}~\mathop \sum\limits_{t=1}^{L} \mathbb{E}\{\Phi_{1,t}+\Phi_{2,t}\}  \\
s.t.&~\eqref{f_1}-\eqref{f_6},
\end{align}
\end{subequations}
where $\mathbb{E}$ denotes the expectation operator, which acts on random system parameters, e.g., electricity price, outdoor temperature, and number of occupants. Decision variables of \textbf{P1} are $m_{i,t}$ and $\sigma_t$.

It is very challenging to solve \textbf{P1} due to the following reasons. Firstly, the thermal dynamics model of $T_{i,t}$ is unknown. Secondly, there are temporally and spatially coupled constraints about indoor temperature $T_{i,t}$ and $\text{CO}_2$ concentration $O_{i,t}$. For example, indoor temperature in a zone at time slot $t+1$ depends on its indoor temperature at time slot $t$ and indoor temperatures in adjacent zones. Similarly, $\text{CO}_2$ concentration at time slot $t+1$ depends on $\text{CO}_2$ concentration in all zones at time slot $t$ according to the expression of $O_t^{\text{mix}}$. Thirdly, the discrete solution space in each time slot is extremely large. For example, when $M=10$, $N=50$, $Z=10$, the total number of solutions is $10^{51}$. Fourthly, control decisions for different zones are coupled by a non-convex and non-separable objective function. Lastly, there are some uncertain system parameters, e.g., electricity price, outdoor temperature, and number of occupants. When taking the above-mentioned challenges into consideration, existing model-based and model-free approaches about building energy optimization are not applicable to \textbf{P1}. For example, DDPG-based HVAC control method in \cite{YuIoT2019} is incapable of dealing with discrete variables in \textbf{P1}, while DQN-based HVAC control method in \cite{Wei2017} is not scalable to the number of zones. Although independent DQN is adopted for each zone, heat transfer among adjacent zones is neglected. Moreover, the way of designing reward function in \cite{Wei2017} is not applicable to the optimization problem \textbf{P1}, which considers the adjustment of AHU damper position.

\subsection{Energy Cost Problem Reformulation}\label{s2_4}
To address the above challenges, we are motivated to design a scalable DRL-based HVAC control algorithm. To be specific, we first reformulate \textbf{P1} as a Markov game\cite{Littman1994}, which is a multi-agent extension of Markov decision process. Then, we propose a model-free control algorithm to solve the Markov game in Section~\ref{s3} based on multi-agent DRL. Since there are $N+1$ variables in \textbf{P1}, $N+1$ agents are considered in the Markov game related to \textbf{P1}. Specifically, a Markov game can be defined by a set of states, $S$, a collection of action sets (each action set is associated with each agent in the environment), $A_1$,~$\cdots$,~$A_{N+1}$, a state transition function, $F:~S\times A_1\times \ldots \times A_{N+1}\rightarrow \Pi(S)$, which defines the probability distribution over possible next states, given the current state and actions for all agents, and a reward function for each agent $i$ ($1\leq i \leq N+1$), $R_i:~S\times A_1\times \ldots \times A_{N+1}\rightarrow \mathbb{R}$. In a Markov game, each agent $i$ takes action $a_i\in A_i$ based on its local observation $o_i\in \mathcal{O}_i$, where $o_i$ contains partial information of the global state $s\in S$. The aim of the agent $i$ is to maximize its expected return by learning a policy $\pi_i:~\mathcal{O}_i\rightarrow \Pi(A_i)$, which maps the agent's local observation $o_i\in \mathcal{O}_i$ into a distribution over its set of actions. Here, the return is the sum of discounted rewards received over the future, i.e., $\sum\nolimits_{j=0}^{\infty}\gamma^j r_{i,t+j+1}(s_t,a_{1,t},\cdots,a_{N+1,t})$, where $\gamma\in[0,1]$ is a discount factor and $r_{i,t+1}\in R_i$ is the reward received by the agent $i$ at slot $t$. Since multi-agent DRL used in Section~\ref{s3} does not require the information of state transition function, we mainly focus on designing three components of the Markov game associated with \textbf{P1}, i.e., state, action, and reward function.

\subsubsection{State}
In zone $i$ ($1\leq i \leq N$), agent $i$ takes action based on its local observation $o_i$ so that indoor temperature and $\text{CO}_2$ concentration can be maintained within the comfortable range. Since indoor temperature $T_{i,t}$ is related to outdoor temperature $T_t^{\text{out}}$ and indoor temperatures in adjacent zones $T_{z,t}$ ($z\in \mathcal{N}_i$), $T_t^{\text{out}}$ and neighbor information should be selected as the parts of system state. Moreover, $\text{CO}_2$ concentration $O_{i,t}$ is associated with the number of occupants $K_{i,t}$. In addition, the common aim of all agents is to minimize HVAC energy cost, which is related to electricity price $\lambda_t$. Based on the above analysis, the local observation of agent $i$ at slot $t$ is designed as follows: $o_{i,t}=(T_t^{\text{out}},T_{i,t},T_{z,t}|_{\forall z\in \mathcal{N}_i},\lambda_t,t',K_{i,t},O_{i,t})$, where $t'$ denotes the time slot index in a day, i.e., $t'=\mod(t,96)$ when $\tau=15$ minutes. Since the action of agent $N+1$ has a large impact on $\text{CO}_2$ concentration and HVAC energy cost, $o_{N+1}$ can be described by $o_{N+1,t}=(\lambda_t,t',K_{1,t},\cdots,K_{N,t},O_{1,t},\cdots,O_{N,t})$. Taking local observations of all agents at slot $t$ into consideration, we have $o_t=(o_{1,t},\cdots,o_{N+1,t})$. For simplicity, the global state $s_t$ is selected to be $o_t$.

\subsubsection{Action}
For agent $i$ ($1\leq i \leq N$), its action is $m_{i,t}$. For agent $N+1$, its action is $\sigma_t$. Thus, the action of all agents can be written as $a_t=(m_{1,t},m_{2,t},\cdots,m_{N,t},\sigma_{t})$.

\subsubsection{Reward function}
According to the theory of Markov decision process, the transition of the environment state from $s_{t-1}$ to $s_t$ could be triggered by the execution of $a_{t-1}$. Then, the reward $r_t$ will be returned. To describe the system transition more concisely, a 4-tuple is used, i.e., $(s_{t-1},a_{t-1},s_t,r_t)$. Under the Markov game environment, $r_t=(r_{1,t},\cdots,r_{i,t},\cdots,r_{N+1,t})$. Since our aim is to minimize the HVAC energy cost while maintaining the comfortable temperature and $\text{CO}_2$ concentration range in all zones, the reward function of an agent is related to four parts, i.e., the penalty for the energy consumption of the supply fan, the penalty for the energy consumption of the cooling coil, the penalty for zone temperature deviation, and the penalty for zone $\text{CO}_2$ concentration violation.

Note that the penalty for the energy consumption of the supply fan is related to $(\sum\nolimits_{i=1}^N m_{i,t})^3$, which is a non-separable function over $m_{i,t}$. To enable the coordination among different zones, the penalty for the energy consumption of the supply fan should be imposed on $N$ zone agents. Since the action of agent $N+1$ has no impact on the energy consumption of the supply fan, we have $r_{N+1,1,t}=0$. For zone agents, the first part of $r_{i,1,t}$ ($1\leq i\leq N$) is designed as follows,
\begin{align} \label{f_12}
r_{i,1,t}(s_{t-1},a_{t-1})=-\frac{\Phi_{1,t-1}}{N},~\forall~t.
\end{align}

According to \eqref{f_8}, the energy consumption of the cooling coil depends on $m_{i,t}$ ($1\leq i\leq N$) and $\sigma_t$. Thus, the penalty for the energy consumption of the cooling coil should be imposed on $N+1$ agents. Without loss of generality, $r_{N+1,2,t}(s_{t-1},a_{t-1})=-\frac{p_{t-1}\lambda_{t-1}\tau}{N+1}$. For zone agents, the second part of $r_{i,2,t}$ is designed by
\begin{align} \label{f_13}
r_{i,2,t}(s_{t-1},a_{t-1})=-\frac{N}{N+1} p_{i,t-1}\lambda_{t-1}\tau,~\forall~t.
\end{align}

To maintain the comfortable temperature range at slot $t$, an action should be taken by each zone agent at slot $t-1$. Then, the reward related to zone temperature deviation is given by
\begin{equation}\label{f_14}
r_{i,3,t}(s_t)=-I_{i,t}({\left[ {{T_{i,t}} - {T_i^{\max }}} \right]^ + } + {\left[ {{T_i^{\min}} - {T_{i,t}}} \right]^+}),
\end{equation}
which means that $r_{i,3,t}(s_t)=0$ if $T_i^{\min}\leq T_{i,t}\leq T_i^{\max}$ and $K_{i,t}>0$. Otherwise, $r_{i,3,t}(s_t)=-(T_{i,t}-T_i^{\max})$ if $T_{i,t}>T_i^{\max}$ and $K_{i,t}>0$, and $r_{i,3,t}(s_t)=-(T_i^{\min}-T_{i,t})$ if $T_{i,t}<T_i^{\min}$ and $K_{i,t}>0$. Note that $I_{i,t}=0$ if $K_{i,t}=0$, and $I_{i,t}=1$ if $K_{i,t}>0$. Therefore, $r_{i,3,t}(s_t)=0$ if there is no any occupant in zone $i$ at slot $t$. Since zone temperature deviation is not relevant to the action of agent $N+1$, we have $r_{N+1,3,t}=0$.

According to \eqref{f_3}, the zone $\text{CO}_2$ concentration violation depends on $m_{i,t}$ ($1\leq i\leq N$) and $\sigma_t$. Thus, the penalty for zone $\text{CO}_2$ concentration violation should be imposed on $N+1$ agents. Without loss of generality, $r_{N+1,4,t}(s_{t})=-\frac{1}{N+1}\sum\nolimits_{i=1}^N{(I_{i,t}\left[{{O_{i,t}}-{O_i^{\max }}} \right]^+)}$. For all zone agents, the fourth part of $r_{i,4,t}$ can be designed by
\begin{align} \label{f_15}
r_{i,4,t}(s_{t})=-\frac{N}{N+1}{(I_{i,t}\left[{{O_{i,t}}-{O_i^{\max }}} \right]^+)},~\forall~t.
\end{align}

Taking four parts into consideration, the reward function of agent $i$ equals $\alpha(r_{i,1,t}(s_{t-1},a_{t-1})+r_{i,2,t}(s_{t-1},a_{t-1}))+\beta r_{i,4,t}(s_{t})+r_{i,3,t}(s_t)$, where $\alpha$ and $\beta$ are positive weight coefficients in $^oC/\$$ and $^oC/\text{ppm}$, respectively.

\begin{figure}[!htb]
\centering
\includegraphics[scale=0.63]{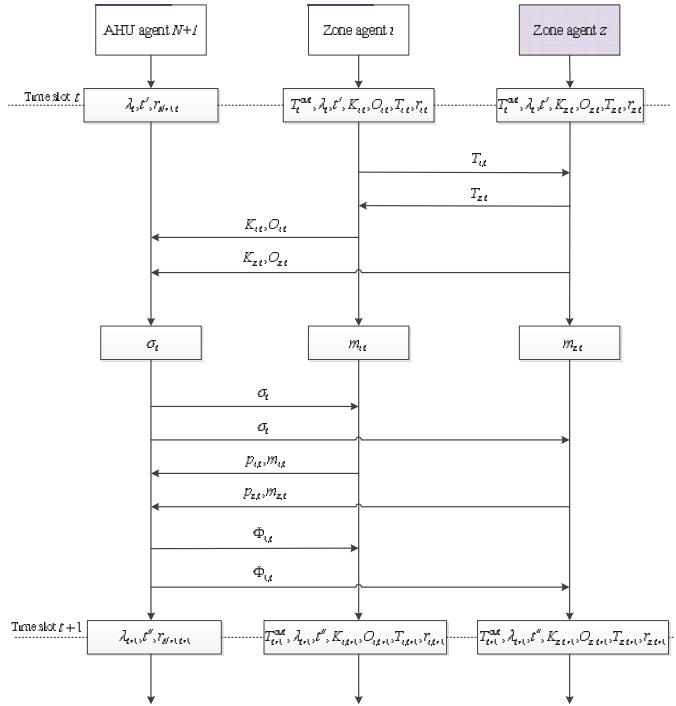}
\caption{Information exchange among different agents.}\label{fig_2}
\end{figure}

To obtain the information related to state and reward function, agents have to coordinate with each other. To be specific, the necessary information in system state and reward function should be collected via information exchanges among different agents as shown in Fig.~\ref{fig_2}. After the state information is obtained, action $a_t=(m_{1,t},~m_{2,t},\cdots,~m_{N,t},~\sigma_t)$ is taken. Then, the new state is observed at beginning of time slot $t+1$ and the reward function $r_{i,t}$ is calculated. Note that number of occupants in each zone can be measured by doorway electronic counting sensors, indoor temperature in each zone and outdoor temperature can be obtained by temperature sensors. In addition, electricity price can be accessed from the website of local electric company.

\begin{figure*}[!htb]
\centering
\includegraphics[scale=0.85]{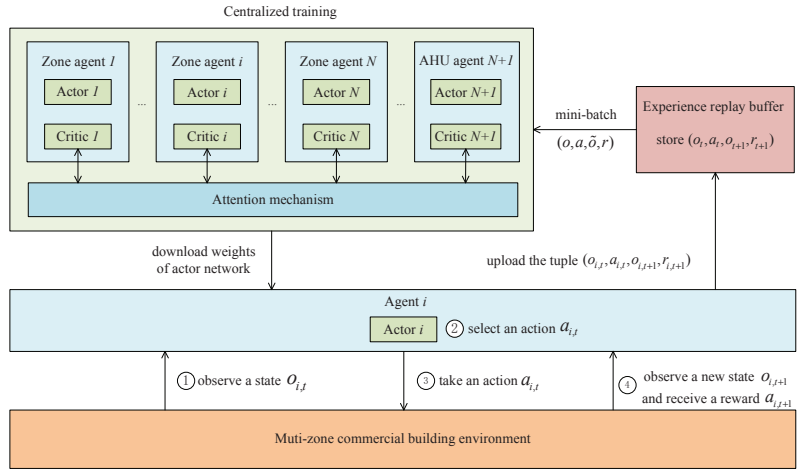}
\caption{The architecture for the proposed MADRL-based HVAC control algorithm.}\label{fig_3}
\end{figure*}

\section{MADRL-based HVAC Control Algorithm}\label{s3}
To solve the Markov game in Section~\ref{s2_4}, we design a MADRL-based HVAC control algorithm. Specifically, multi-actor-attention-critic (MAAC) approach\cite{Iqbal2019} for MADRL is adopted. In the following parts, we first introduce the basic principle of MAAC approach. Then, we describe the proposed HVAC control algorithm.

\subsection{Basic Principle of MAAC Approach}
MAAC approach is a kind of actor-critic methods, which learn approximations to both policy and value functions. Typically, action-value function $Q(s_t,a_t)$ is used and it represents the expected return of taking action $a_t$ in state $s_t$, i.e., $Q(s_t,a_t)=\mathbb{E}\{\sum\nolimits_{j=0}^{\infty}\gamma^j r_{t+j+1}(s_t,a_t)\}$. The approximation of action-value function (i,e., $Q_{\psi}(s_t,a_t)$ and $\psi$ denotes the weight parameter of critic network) can be learned through off-policy temporal-difference learning by minimizing the following loss function, i.e.,
\begin{equation}\label{f_16}
\mathcal{L}_Q(\psi)=\mathbb{E}_{(s,a,\tilde{s},r)\sim \mathcal{D}}[(Q_{\psi}(s,a)-y)^2],
\end{equation}
where $y=r(s,a)+\gamma \mathbb{E}_{\tilde{a}\in \pi(\tilde{s})}[Q_{\bar{\psi}}(\tilde{s},\tilde{a})]$, $Q_{\bar{\psi}}$ is the target value function, and $\mathcal{D}$ is an experience replay buffer that stores past system transitions $(s,a,\tilde{s},r)$.

Similarly, the approximation of policy function $\pi_{\theta}(a|s)$ (note that $\theta$ denotes the weight parameter of actor network) can be learned by policy gradient methods, and the policy gradient can be written as follows,
\begin{equation}\label{f_17}
\nabla_{\theta}J(\theta)=\mathbb{E}_{s\thicksim \mathcal{D}, a\thicksim \pi }[\nabla_{\theta}\text{log}(\pi_{\theta}(a|s))Q_{\psi}(s,a)].
\end{equation}

To encourage exploration and avoid convergence to non-optimal deterministic policies, a soft actor-critic method is adopted in MAAC approach by incorporating an entropy term as follows,
\begin{equation}\label{f_18}
\nabla_{\theta}J(\theta)=\mathbb{E}_{s\thicksim \mathcal{D}, a\thicksim \pi}[\nabla_{\theta}\text{log}(\pi_{\theta}(a|s))\rho(s,a)],
\end{equation}
where $\rho(s,a)=-\varphi\text{log}(\pi_{\theta}(a|s))+ Q_{\psi}(s,a)-b(s)$, $\varphi$ is the temperature parameter determining the balance between maximizing entropy and reward, $b(s)$ is a state-dependent baseline. Accordingly, the target value $y$ used in the loss function for estimating $Q_{\psi}(s,a)$ can be revised as follows,
\begin{equation}\label{f_19}
y=r(s,a)+\gamma \mathbb{E}_{\tilde{a}\in \pi(\tilde{s})}[-\varphi\text{log}(\pi_{\bar{\theta}}(\tilde{a}|\tilde{s}))+Q_{\bar{\psi}}(\tilde{s},\tilde{a})],
\end{equation}
where $\pi_{\bar{\theta}}(\tilde{a}|\tilde{s})$ denotes the target policy function with parameter $\bar{\theta}$.

In addition to the above-mentioned soft actor-critic method, MAAC approach adopts attention mechanism, which can learn the critic for each agent by selectively paying attention to the information from other agents. In other words, the contributions from other agents are considered when calculating the action-value function $Q_i^{\psi}(o,a)$ for agent $i$. To be specific, $Q_i^{\psi}(o,a)$ for agent $i$ can be calculated as follows,
\begin{equation}\label{f_20}
Q_i^{\psi}(o,a)=f_i(g_i(o_i),x_i),
\end{equation}
where $f_i$ is a two-layer multi-layer perception (MLP), $g_i$ is a one-layer MLP embedding function, and $x_i$ denotes the total contribution from other agents.


Let $q_i$ be a one-layer MLP embedding function and $e_i=q_i(o_i,a_i)$. Then, we have
\begin{equation}\label{f_21}
x_i=\sum\nolimits_{j\neq i}\omega_jh(W_ve_j),
\end{equation}
where $W_v$ is a shared matrix that transforms $e_j$ into a ``value", $h$ is a non-linear activation function, $\omega_j$ is the attention weight associated with agent $j$ and can be obtained as follows,
\begin{equation}\label{f_22}
\omega_j=\text{exp}^{((W_ke_j)^TW_qe_i)}/\sum\nolimits_{j=1}^{N}\text{exp}^{((W_ke_j)^TW_qe_i)},~\forall~j,
\end{equation}
where $W_k$ and $W_q$ are shared matrixes that transform $e_j$ into a ``key" and transform $e_i$ into a ``query" respectively. Note that the above-mentioned ``value", ``key" and ``query" are similar to those in the key-value memory model\cite{Oh2016}.

Due to the sharing of three learnable parameters (i.e., $W_k, W_q, W_v$), all critics are updated to minimize a joint regression loss function as follows,
\begin{equation}\label{f_23}
\mathcal{L}_Q(\psi)=\sum\limits_{i=1}^{N+1}\mathbb{E}_{(o,a,\tilde{o},r)\sim \mathcal{D}}[(Q_i^{\psi}(s,a)-y_i)^2],
\end{equation}
where $y_i=r_i(o,a)+\gamma \mathbb{E}_{\tilde{a}\in \pi_{\bar{\theta}}(\tilde{o})}[-\varphi\text{log}(\pi_{\bar{\theta_i}}(\tilde{a}_i|\tilde{o}_i))+Q_i^{\bar{\psi}}(\tilde{o},\tilde{a})]$. Similarly, the gradient used for updating individual policies is given by
\begin{equation}\label{f_24}
\nabla_{\theta_i}J(\theta)=\mathbb{E}_{o\thicksim \mathcal{D}, a\thicksim \pi}[\nabla_{\theta_i}\text{log}(\pi_{\theta_i}(a_i|o_i))\rho_i(o_i,a_i)],
\end{equation}
where $\rho_i(o_i,a_i)=-\varphi\text{log}(\pi_{\theta_i}(a_i|o_i))+ Q_i^{\psi}(o,a)-b(o,a_{\backslash i})$, $\backslash i$ denotes the set of agents except $i$. Here, $Q_i^{\psi}(o,a)-b(o,a_{\backslash i})$ is called as the multi-agent advantage function, which can indicate that whether the current action would cause an increase in expected return, where $b(o,a_{\backslash i})=\sum\nolimits_{\tilde{a}_i\in A_i}\pi_{\bar{\theta}_i}(\tilde{a}_i|o_i)Q_i^{\psi}(o,(\tilde{a}_i,a_{\backslash i}))$.

\begin{algorithm}[h]
\caption{Training Algorithm}
\label{alg_1}
\setcounter{AlgoLine}{0}
\LinesNumbered
\KwIn{The traces of electricity price, outdoor temperature and number of occupants}
\KwOut{The weights of actor network and critic network, i.e., $\theta$ and $\psi$}

Initialize the capacity of experience replay buffer $\mathcal{D}$\;

Initialize environments with $N+1$ agents\;

Initialize the weights of target networks $Q_i^{\bar{\psi}}$ and $\pi_i^{\bar{\theta}}$ by copying: $\bar{\psi}\Leftarrow\psi$, $\bar{\theta}\Leftarrow\theta$\;

\For{episode=1,~2,~$\cdots$,~$Y$}
{
   Reset environments, and get initial observation state $o_{i,1}$ for each agent $i$\;

  \For{$t$=1,~2,~$\cdots$,~$P$}
  {
    Select actions $a_{i,t}\sim \pi_i^{\theta}(\cdot|o_{i,t})$ for each agent $i$\;

    Send actions $a_{i,t}$ to all parallel environments and get $o_{i,t+1}$ and $r_{i,t+1}$\;

    Store transitions $(o_{t},~a_{t},~o_{t+1},~r_{t+1})$ in $\mathcal{D}$\;

    \If{$G_{\text{memory}}\geq B_{\text{size}}$ and \text{mod}($t$,$T_{\text{update}}$)=0}
      {

    Sample mini-batch $\mathbf{B}$ with $B_{\text{size}}$ transitions ($o,a,\tilde{o},r$) from $\mathcal{D}$\;

    Calculate $Q_i^{\psi}(o_{i}^l,a_{i}^l)$ for all $i$ in parallel, $1\leq l\leq B_{\text{size}}$, where $a_i^l$ and $o_{i}^l$ denotes $l$th $a_i$ and $o_i$ in $\mathbf{B}$\;

    Calculate $\tilde{a}_{i}^l\sim \pi_i^{\bar{\theta}}(\tilde{o}_{i}^l)$ for all $i$ and $l$\;

    Calculate $Q_i^{\bar{\psi}}(\tilde{o}_{i}^l,\tilde{a}_{i}^l)$ for all $i$ and $l$\;

    Update critic network by minimizing the joint regressive loss function \eqref{f_23}\;

    Calculate $a_{i}^l \sim \pi_i^{\bar{\theta}}(o_{i}^l)$ for all $i$ in parallel\;

    Calculate $Q_i^{\psi}(o_{i}^l,~a_{i}^l)$ for all $l$ and $i$\;

    Update policies using \eqref{f_24}\;

    Update the weights of target networks:\;

    $\bar{\psi }\leftarrow \xi \psi+(1-\xi)\bar{\psi}$,~$\bar{\theta} \leftarrow \xi \theta+(1-\xi)\bar{\theta}$\;
    }

  }
}
\end{algorithm}

\begin{algorithm}[h]
\caption{Execution Algorithm}
\label{alg_2}
\setcounter{AlgoLine}{0}
\LinesNumbered
\KwIn{The weights of the actor network, i.e., $\theta$}
\KwOut{Action $a_{t}$}

All agents receive initial local observation $o_{1}=(o_{1,1},\cdots,o_{N+1,1})$\;

\For{$t$=1,~2,~$\cdots$,~$H_{\text{test}}$}
{

Each agent $i$ selects its action $a_{i,t}$ in parallel according to the learned policy $\pi_{\theta}(\cdot|o_{i,t})$ at the beginning of slot $t$\;

Each agent $i$ takes action $a_{i,t}$ in parallel, which affects the operation the HVAC system\;

Each agent $i$ receives new observation $o_{i,t+1}$ at the end of slot $t$\;

}
\end{algorithm}

\subsection{The Proposed MADRL-based HVAC Control Algorithm}
The proposed MADRL-based HVAC control algorithm consists of two parts, i.e., training algorithm and execution algorithm, which can be illustrated by Algorithm 1 and Algorithm 2 respectively. In Algorithm 1, agent $i$ interacts with multi-zone commercial building environment at each time slot $t$. Following the procedure in Fig.~\ref{fig_2}, the transition tuple $(o_{i,t},a_{i,t},o_{i,t+1},r_{i,t+1})$ of each agent is collected and stored in experience replay buffer $\mathcal{D}$, which can be shown in lines 7-9 of Algorithm 1. When the length of the buffer $G_{\text{memory}}$ is larger than batch size $B_{\text{size}}$, a mini-batch data would be randomly sampled from the buffer $\mathcal{D}$ every $T_{\text{update}}$ slot and used to train critic/actor networks. Note that the above storing and sampling technique is called as experience replay, which has many advantages\cite{Mnih2015}, e.g., higher data efficiency, reduced variance of updates, and stronger stability. Lines 12-15 in Algorithm 1 are associated with the update of critic network and Lines 16-18 are related to the update of actor network. Finally, weights of target network are updated in Line 20. To illustrate the above training process, the architecture of the proposed MADRL-based HVAC control algorithm is shown in Fig.~\ref{fig_3}.

After the completion of training process, the weights of actor networks are not updated. Then, the learned policies can be used for practical testing as shown in Algorithm 2. To be specific, each actor takes action $a_{i,t}$ based on the current local observation $o_{i,t}$. Next, actions of actors in all agents are executed by the HVAC system. At the end of time slot $t$, local observation at next time slot $o_{i,t+1}$ is obtained by agent $i$. The above-mentioned procedure repeats until the end of testing stage. Since just the current observation $o_t$ is used for making decisions, the proposed HVAC control algorithm does not require any prior knowledge of uncertain system parameters and building thermal dynamics models. Note that all actors have the same network architecture, which consists of one input layer, multiple hidden layers with Leaky ReLU activation functions, and one output layer with softmax activation function. Similarly, all critics have the same network architecture, which is composed of one input layer, multiple hidden layers with Leaky ReLU activation functions, and one output layer with linear activation function. Since the interaction processes between agents and real building environments may last for a long time in practice\cite{Chen2019}\cite{Zou2019}, some approaches have been proposed to reduce the ``real environment" dependency, e.g., pre-training\cite{Chen2019}, building energy model calibration\cite{Zhang2019}, HVAC operation approximation\cite{Zou2019}. For example, EnergyPlus can be used to create building energy models. Then, these models are calibrated with several months of actual observable data (e.g., weather condition, energy consumption). Next, the calibrated models are adopted as environment simulators for the training of DRL agents\cite{Zhang2019}. Although some advances have been made in \cite{Zhang2019}, the DRL agent's exposure to real-world HVAC operational data is limited since just three months of the observed data are used for calibrating building energy models. To overcome this drawback, several years of historical data can be used to pre-train a control policy. Then, the control policy is improved continually in a real building environment using online learning algorithm\cite{Chen2019}. In addition, the real-world HVAC operations can be approximated based on building historical data by implementing LSTM networks\cite{Zou2019}, which take the current state and action as inputs and predict the next state and reward. Then, the obtained LSTM networks can be used to create training environments for DRL-agents.

\section{Performance Evaluation}\label{s4}

\subsection{Experimental Setup}
\begin{figure*}
\centering
\subfigure[Retail commercial price]{
\begin{minipage}[b]{0.31\textwidth}
\includegraphics[width=1\textwidth]{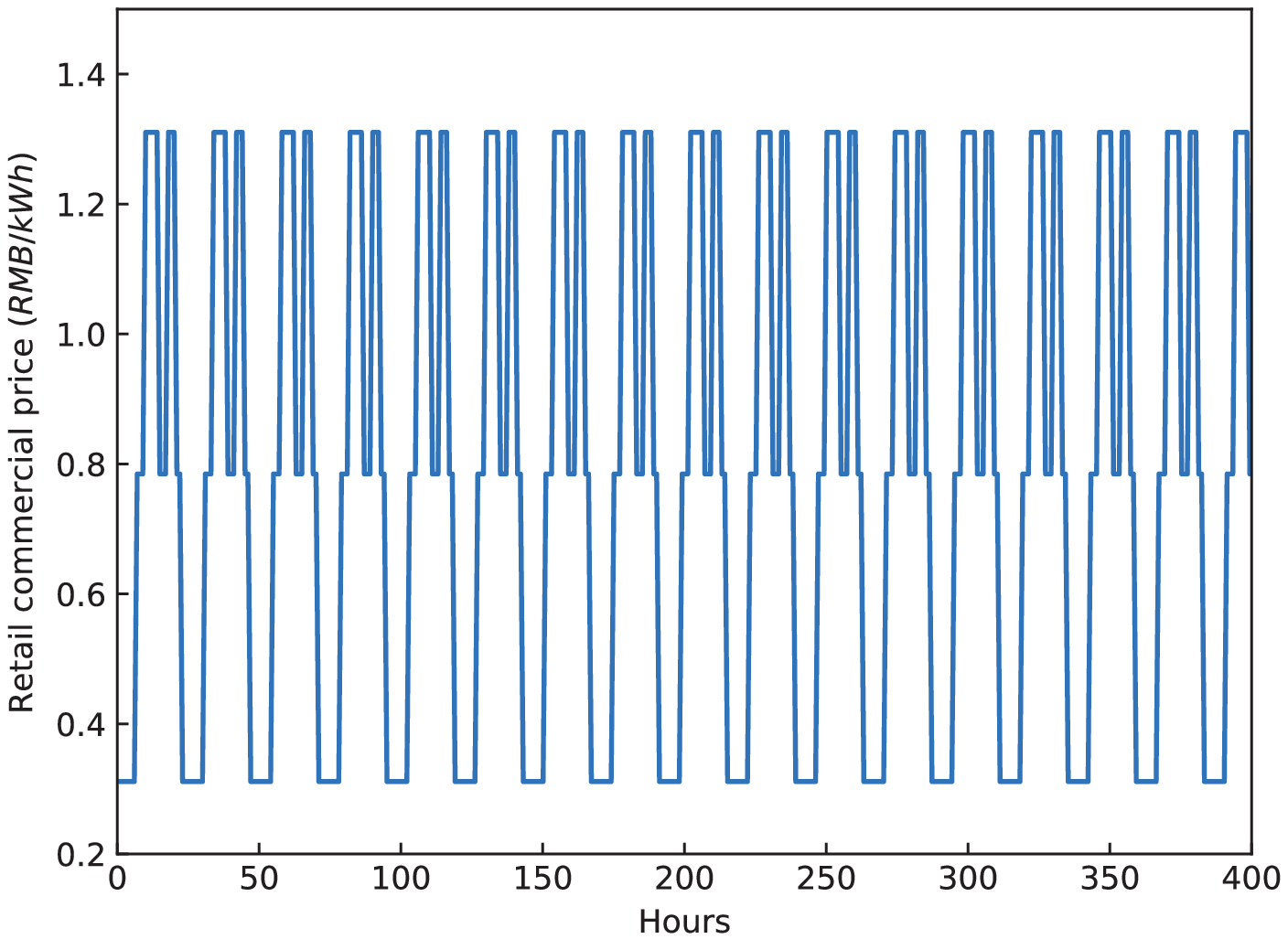}
\end{minipage}
}
\subfigure[Outdoor temperature]{
\begin{minipage}[b]{0.31\textwidth}
\includegraphics[width=1\textwidth]{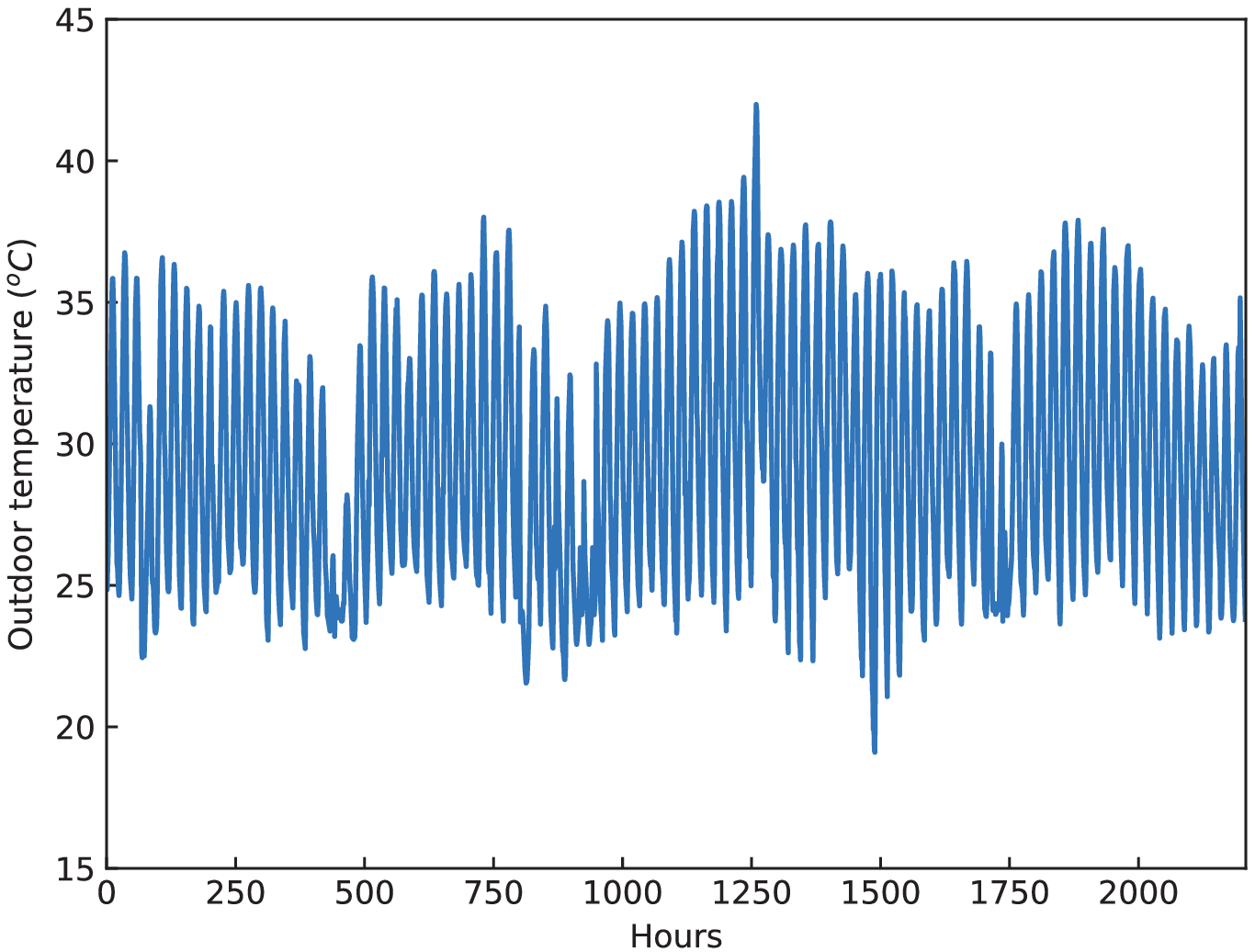}
\end{minipage}
}
\subfigure[Number of occupants]{
\begin{minipage}[b]{0.31\textwidth}
\includegraphics[width=1\textwidth]{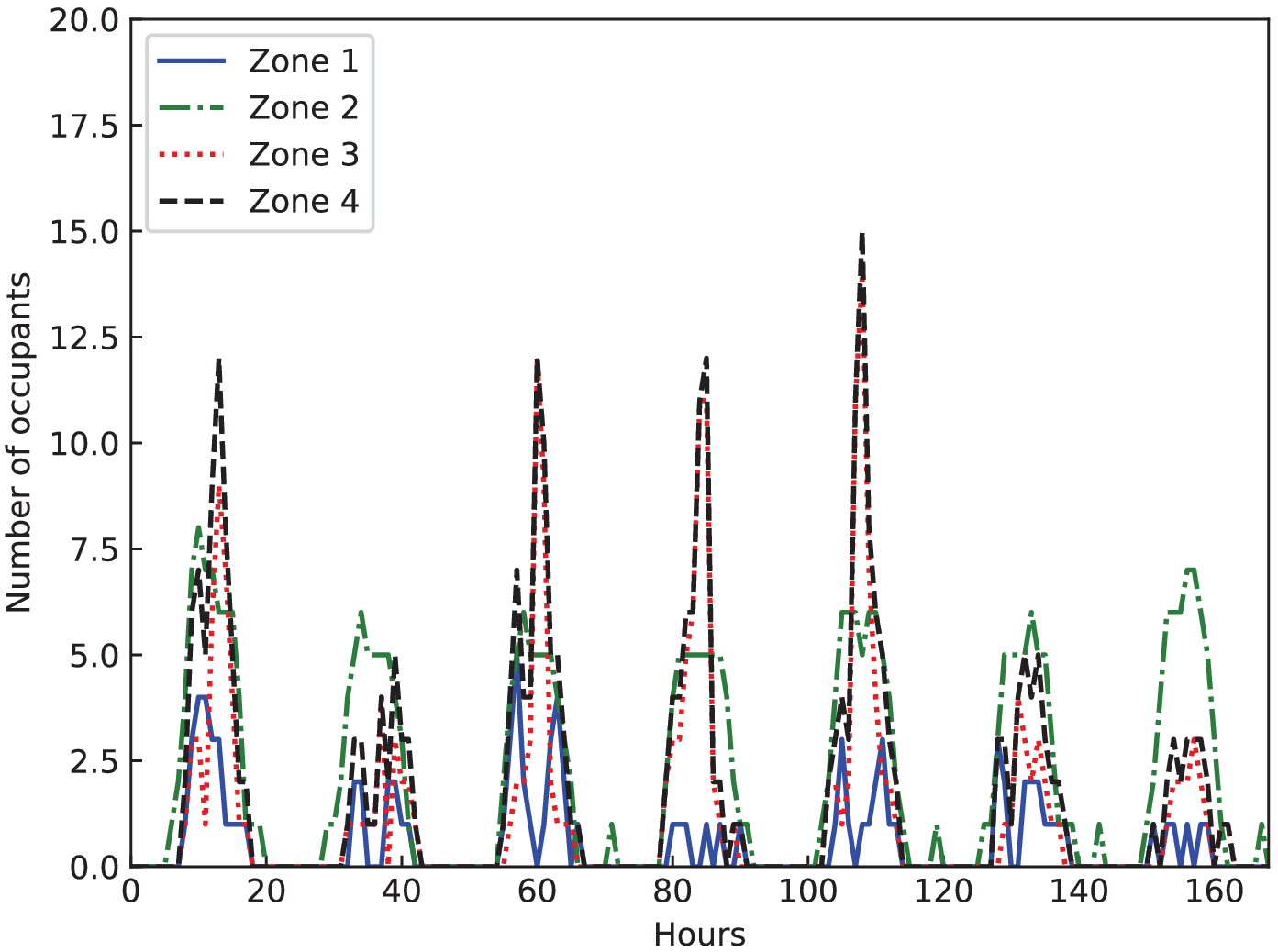}
\end{minipage}
}
\caption{Real-world traces used in simulations.} \label{fig_4}
\end{figure*}

Real-world retail commercial electricity price data\footnote{http://www.beijing.gov.cn/zhengce/gfxwj/201905/t20190531\_82985.html}, outdoor temperature data from Pecan Street database\footnote{https://www.pecanstreet.org/}, and zone occupancy data from \cite{COD2017} are used in simulations. To be specific, Time-of-Use (ToU) commercial price in Beijing and hourly outdoor temperature in Austin, Texas, USA during June 1 to August 31, 2018 are used. Note that ToU price in Beijing is independent of environment conditions, it is reasonable to use real-world traces from different locations. Since just three zone occupancy traces are available in \cite{COD2017}, the occupancy trace in zone 4 is obtained based on that in zone 1 and zone 3. Moreover, we mainly consider the average number of occupants during each hour and reduce the number of occupants in zone 2 by multiplying a coefficient. Note that the curves of above-mentioned traces are plotted in Fig.~\ref{fig_4}. In simulations, the data in June and July are used to train neural network models and the data in August are adopted for performance testing. The training is conducted on a laptop computer with Intel Core$^{\text{TM}}$ i5-7300HQ CPU@2.5GHz and 24 RAM. In addition, the proposed algorithm is implemented based on Python. Main parameter configurations related to model training are shown in TABLE~\ref{table_1}, where $\alpha_a$ and $\alpha_c$ denote learning rates of actor network and critic network, respectively. $N_a$ and $N_c$ denote the number of neurons in two hidden layers of actor network and critic network respectively. It is worth mentioning that the capacity of experience replay buffer $N_{\text{buffer}}$ has a large impact on the convergence of the training algorithm. According to the analysis of simulation results, a larger $N_{\text{buffer}}$ should be selected for a larger number of agents. To simulate the environment in a multi-zone commercial building, we adopt the temperature dynamics model in \cite{LiangTSG2019} for simplicity, where zone temperature disturbances are assumed to be zero. To illustrate that the proposed algorithm is applicable to any building thermal dynamics model, we evaluate the robustness of the proposed algorithm when thermal disturbances are non-zero in the following parts. When the number of zones is larger than 4, model parameters are repeatedly used. Other constant parameters are selected as follows: $\mu=2\times 10^{-6}W/(g/s)^3$\cite{Zhang2017}, $m_{i,t}\in \{0,0.1,\cdots,0.9,1\}*450g/s$\cite{Zhang2017}, $\sigma_t \in \{0,0.1,\cdots,0.9,1\}$, $\eta=0.8879$\cite{Hao2017}, $COP=5.9153$\cite{Hao2017}, $T_i^{\min}=19^oC$\cite{YuIoT2019}, $T_i^{\max}=24^oC$\cite{YuIoT2019}, $O_i^{\max}=1300$ppm\cite{LiangTSG2019}, $H_{\text{test}}=744$~hours, $\tau=15$~min, $P=96$.

\begin{figure*}
\centering
\subfigure[Total energy cost]{
\begin{minipage}[b]{0.31\textwidth}
\includegraphics[width=1\textwidth]{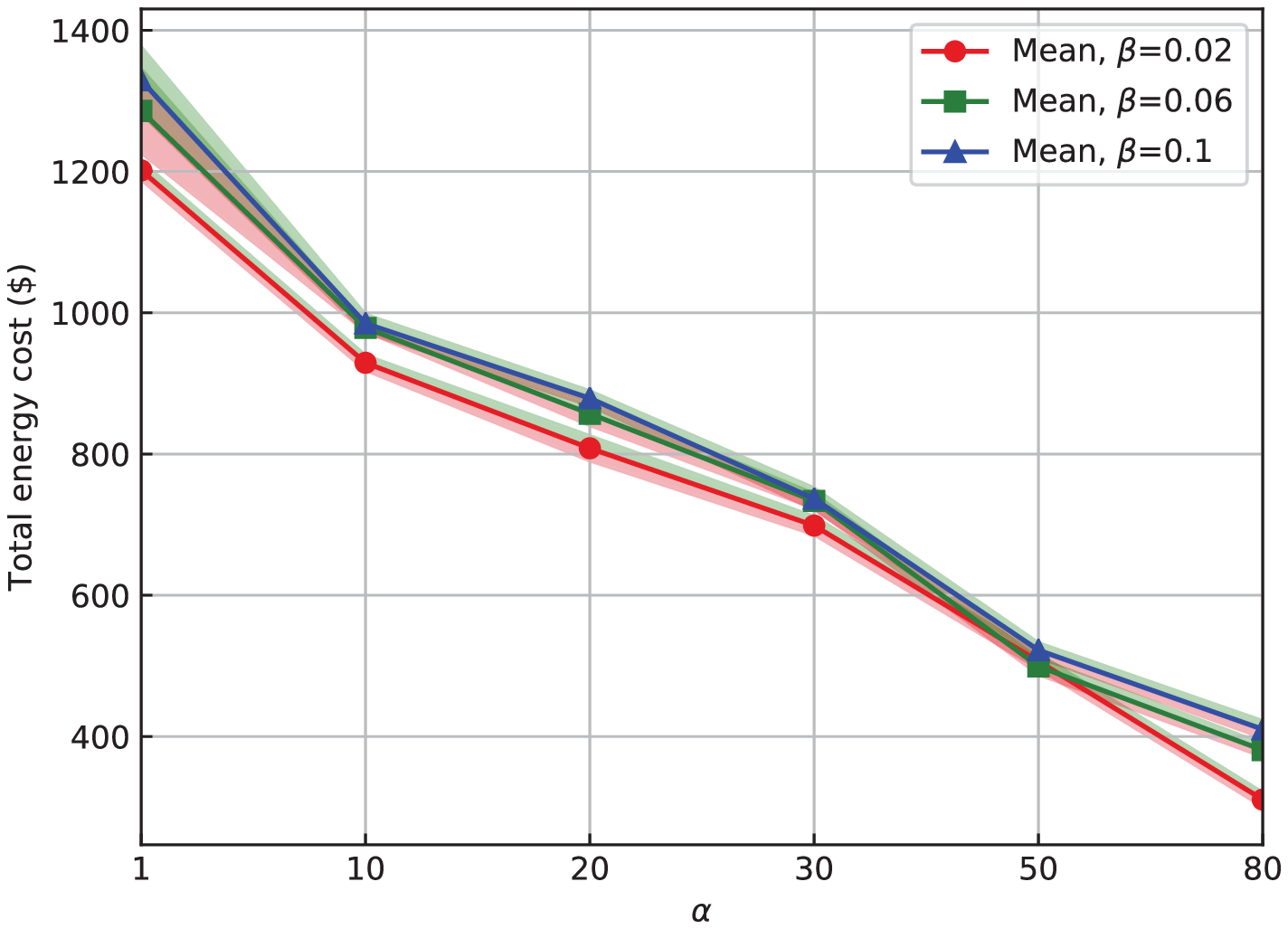}
\end{minipage}
}
\subfigure[ACD ]{
\begin{minipage}[b]{0.31\textwidth}
\includegraphics[width=1\textwidth]{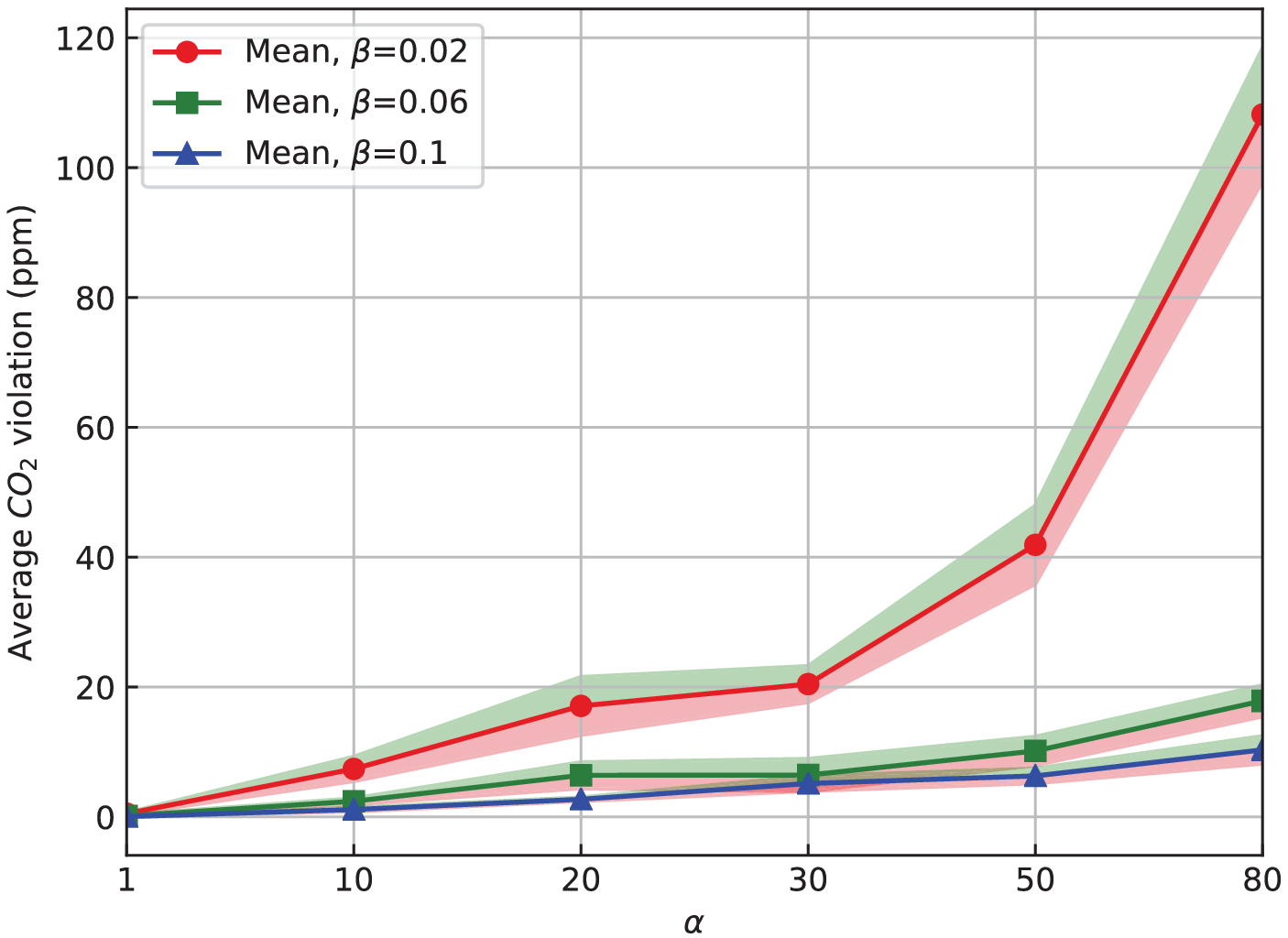}
\end{minipage}
}
\subfigure[ATD]{
\begin{minipage}[b]{0.31\textwidth}
\includegraphics[width=1\textwidth]{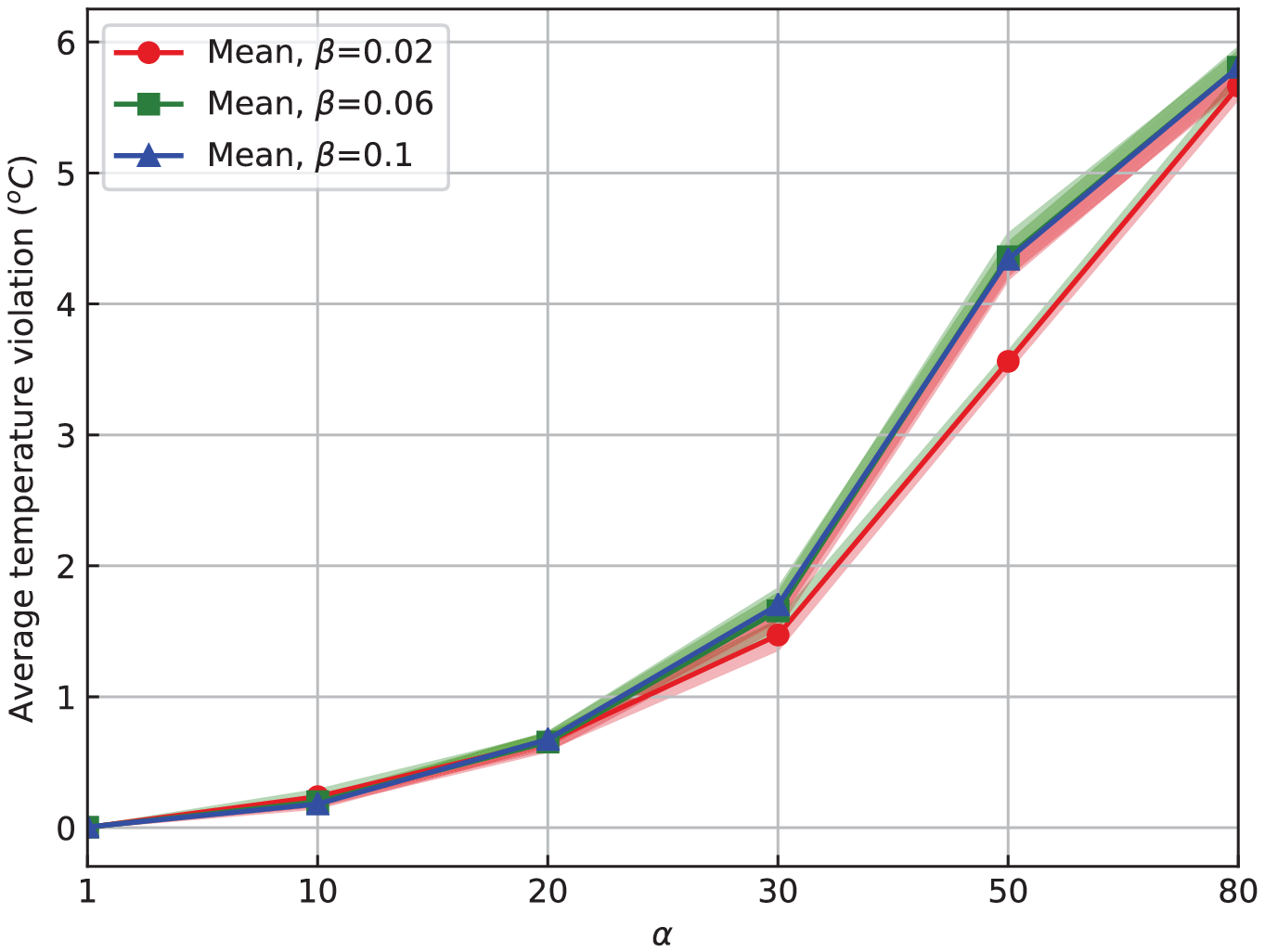}
\end{minipage}
}
\caption{The performance of the proposed algorithm under varying $\alpha$ and $\beta$. Note that 95\% confidence interval across 10 runs is considered.} \label{fig_5}
\end{figure*}

\begin{table}[!htb]
\renewcommand{\arraystretch}{1.3}
\caption{Model training parameters} \label{table_1} \centering
\begin{tabular}{|c|c||c|c||c|c|}
\hline  $Y$                    &5000       &$\alpha_a$            &0.0005     &$\alpha_c$           &0.001    \\
\hline  $\xi$                  &0.001      &$\gamma$               &0.995     &$N_a$                &128      \\
\hline  $N_c$                  &128        &$B_{\text{size}}$      &120       &$\varphi$             &0.1    \\
\hline  $N_{\text{buffer}}$    &4800000    &$T_{\text{update}}$    &1         &Optimizer            &Adam     \\
\hline
\end{tabular}
\end{table}

\subsection{Benchmark Methods}
\begin{enumerate}
  \item \emph{Rule-based Scheme} (\emph{RS}): This scheme adopts ON/OFF policy in \cite{Wei2017} for indoor temperature control. To be specific, $m_{i,t}=m_i^{M}$ if $K_{i,t}>0$ and $T_{i,t}>T_i^{\max}$; $m_{i,t}=m_i^{1}$ if $K_{i,t}>0$ and $T_{i,t}<T_i^{\min}$. For the case that $K_{i,t}>0$ and $T_i^{\min}\leq T_{i,t}\leq T_i^{\max}$, $m_{i,t}$ is not changed. When $K_{i,t}=0$, $m_{i,t}=0$. Moreover, the damper position in the AHU $\sigma_t$ is fixed during the testing period.
  \item \emph{Heuristic Scheme} (\emph{HS}): \emph{HS} uses building thermal dynamics models for building environment control\footnote{Since the proposed MADRL-based algorithm does not require thermal dynamics models of $T_{i,t}$, it is unfair to compare the performance of \emph{HS} with that of the proposed algorithm. However, \emph{HS} can be used to illustrate the necessity of using the proposed algorithm when existing model-based and model-free approaches are not applicable to the original problem \textbf{P1}.}. To be specific, $m_{i,t}=0$ and $\sigma_t=0$ if $K_{i,t}=0$. When $K_{i,t}>0$ and $O_{i,t}+\frac{K_{i,t}\tau\chi}{\vartheta_i}<O_i^{\max}$, $\sigma_{i,t}=1$ and $m_{i,t}=\min\{m_i^{\max},\max\{m_i^{\min},m_{i,1}^{\text{low}}\}\}$, where $m_{i,1}^{\text{low}}$ is the minimum air supply rate of making $T_{i,t}$ below $T_i^{\max}$. When $K_{i,t}>0$ and $O_{i,t}+\frac{K_{i,t}\tau\chi}{\vartheta_i}\geq O_i^{\max}$, $\sigma_{i,t}=\zeta$ and $m_{i,t}=\min\{m_i^{\max},\max\{m_i^{\min},m_{i,2}^{\text{low}}\}\}$, where $m_{i,2}^{\text{low}}=\frac{\kappa\vartheta_i(O_i^{\max}-O_{i,t}-\frac{K_{i,t}\tau\chi}{\vartheta_i})}{\tau((1-\zeta)O_t^{\text{out}}+\zeta O_t^{\max}-O_{i,t})}$ is the minimum air supply rate of making $O_{i,t+1}$ below $O_i^{\max}$, and $O_t^{\max}=\max_i \{O_{i,t}\}$. Finally, $\sigma_t=\frac{1}{N}\sum\nolimits_i\sigma_{i,t}$, which means that $\sigma_t$ is jointly decided by all zones. Since the obtained $m_{i,t}$ and $\sigma_t$ are not necessarily effective decisions in discrete sets, solutions with the nearest distances to them are selected.
\end{enumerate}

\subsection{Comfort-Related Performance Metrics}
To describe the extent of thermal discomfort and indoor air quality discomfort perceived by occupants in all zones at all slots concisely, we adopt two performance metrics, i.e., \emph{Average Temperature Deviation} (\emph{ATD}) and \emph{Average $\text{CO}_2$ Concentration Deviation} (\emph{ACD}). To be specific, $ATD=\frac{1}{N\bar{L}}\sum\nolimits_{i=1}^{N}\sum\nolimits_{t=1}^{L}I_{i,t}({\left[ {{T_{i,t}} - {T_i^{\max }}} \right]^ + } + {\left[ {{T_i^{\min}} - {T_{i,t}}} \right]^+})$, $ACD=\frac{1}{N\bar{L}}\sum\nolimits_{i=1}^{N}\sum\nolimits_{t=1}^{L}I_{i,t}({\left[ {{O_{i,t}} - {O_i^{\max }}} \right]^ + })$, where $\bar{L}=\sum\nolimits_{t=1}^{L}I_{i,t}$ denotes the total number of slots with occupation related to zone $i$.

\subsection{Algorithmic Performance under Varying $\alpha$ and $\beta$}
The performance of the proposed algorithm under varying $\alpha$ and $\beta$ is provided in Fig.~\ref{fig_5}, where the mean value of the total energy cost generally decreases with the increase of $\alpha$ and the decrease of $\beta$. The reason is obvious since $\alpha$ and $\beta$ represent the relative importance weight of energy cost and $\text{CO}_2$ concentration deviation with respect to temperature violation, respectively. Due to the same reason, the mean value of average $\text{CO}_2$ concentration deviation increases given a larger $\alpha$ and a smaller $\beta$ as shown in Fig.~\ref{fig_5}(b). Similarly, given a larger $\alpha$ or a larger $\beta$, the mean value of average temperature deviation increases as shown in Fig.~\ref{fig_5}(c). Therefore, the proposed algorithm can provide a flexible tradeoff between energy cost and comfort. In practice, according to the tolerable $ATD$ and $ACD$, a proper $\alpha$ and $\beta$ could be selected.

\begin{table*}[bth]
\renewcommand{\arraystretch}{1.3}
\caption{$\overline{TEC}$ under different conditions. Note that $\text{N/A}$ denotes that $\overline{TEC}$ is unavailable under the given conditions.} \label{table_2} \centering
\begin{threeparttable}
\begin{tabular}{|c||c|c|c|c|}
\hline  \multicolumn{1}{|c||}{} & $\overline{ATD}\leq 1.2^oC, \overline{ACD}\leq 40\text{ppm}$ & $\overline{ATD}\leq 1^oC, \overline{ACD}\leq 10\text{ppm}$ & $\overline{ATD}\leq 0.01^oC, \overline{ACD}\leq 0.2\text{ppm}$         \\
\hline  \emph{RS}                      &   1726.8671~$\text{RMB}$      & 1726.8671~$\text{RMB}$        & N/A    \\
\hline  \emph{HS}                      &   806.9383~$\text{RMB}$    & N/A                          & N/A    \\
\hline  Proposed                & 764.7521$\pm$10.692~$\text{RMB}$  & 857.0676$\pm$16.5672~$\text{RMB}$ &1285.7955$\pm$59.4744~$\text{RMB}$ \\
\hline
\end{tabular}
\end{threeparttable}
\end{table*}

\begin{figure*}
\centering
\subfigure[$\overline{ATD}$]{
\begin{minipage}[b]{0.31\textwidth}
\includegraphics[width=1\textwidth]{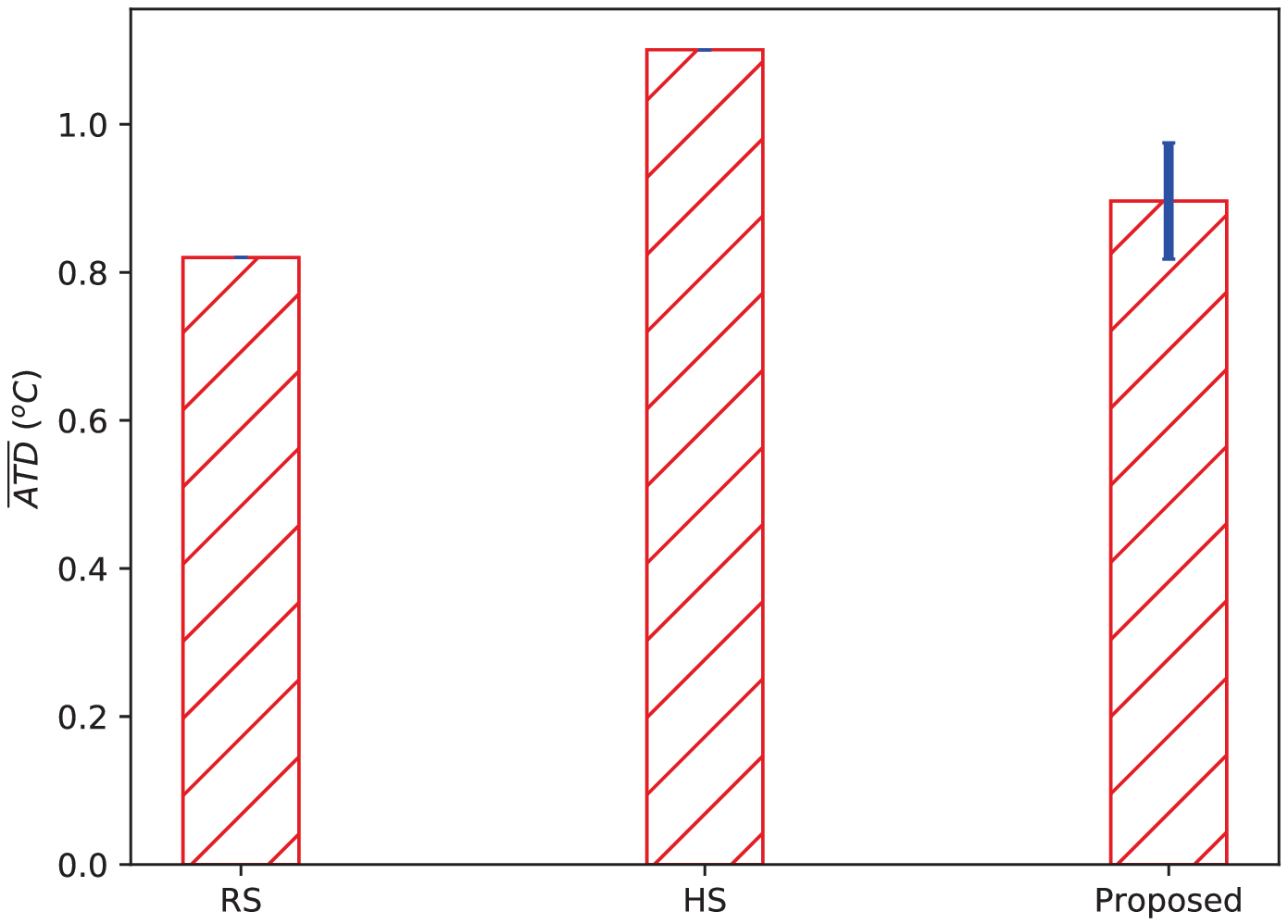}
\end{minipage}
}
\subfigure[$\overline{ACD}$]{
\begin{minipage}[b]{0.31\textwidth}
\includegraphics[width=1\textwidth]{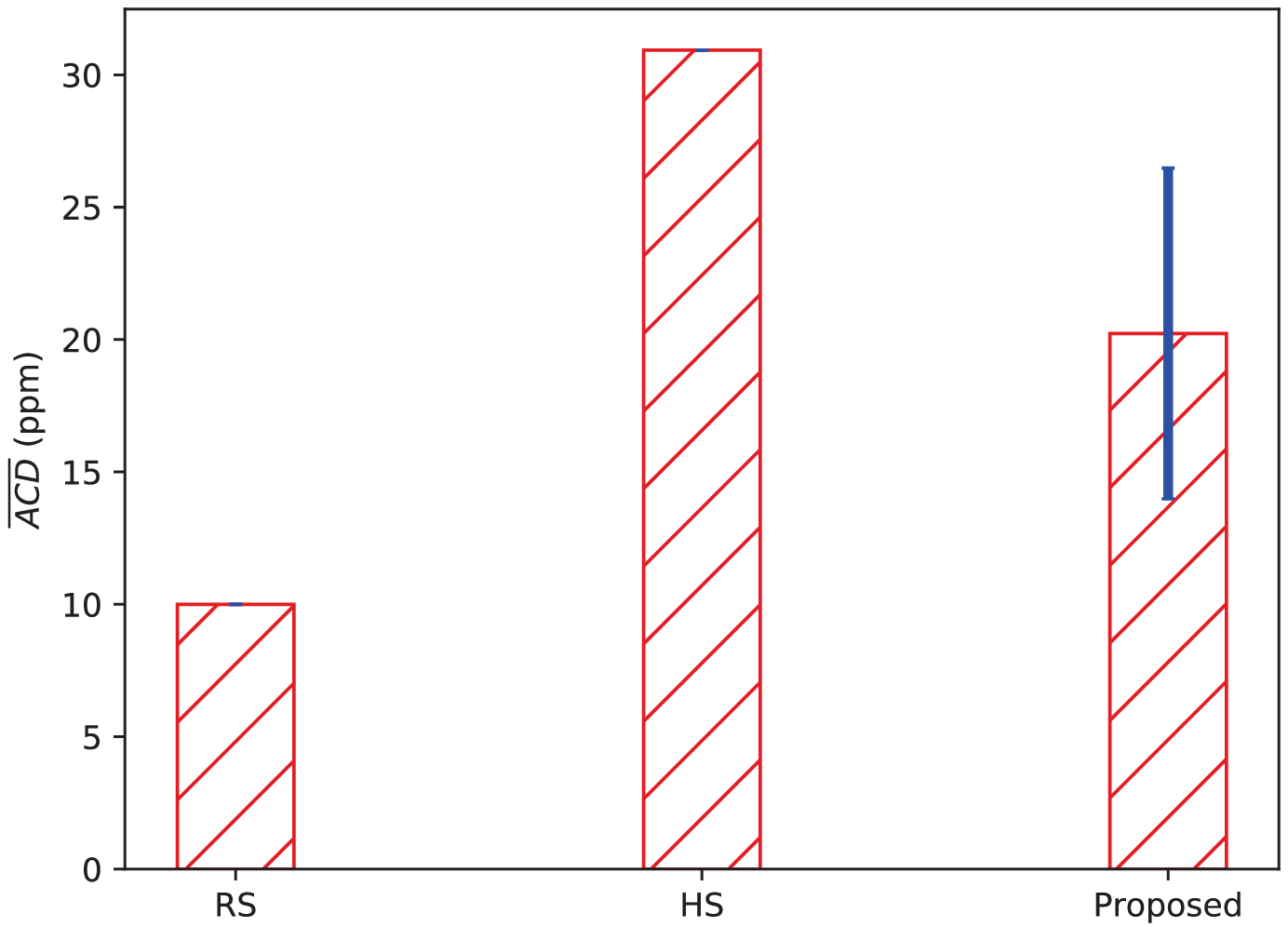}
\end{minipage}
}
\subfigure[$\overline{TEC}$]{
\begin{minipage}[b]{0.31\textwidth}
\includegraphics[width=1\textwidth]{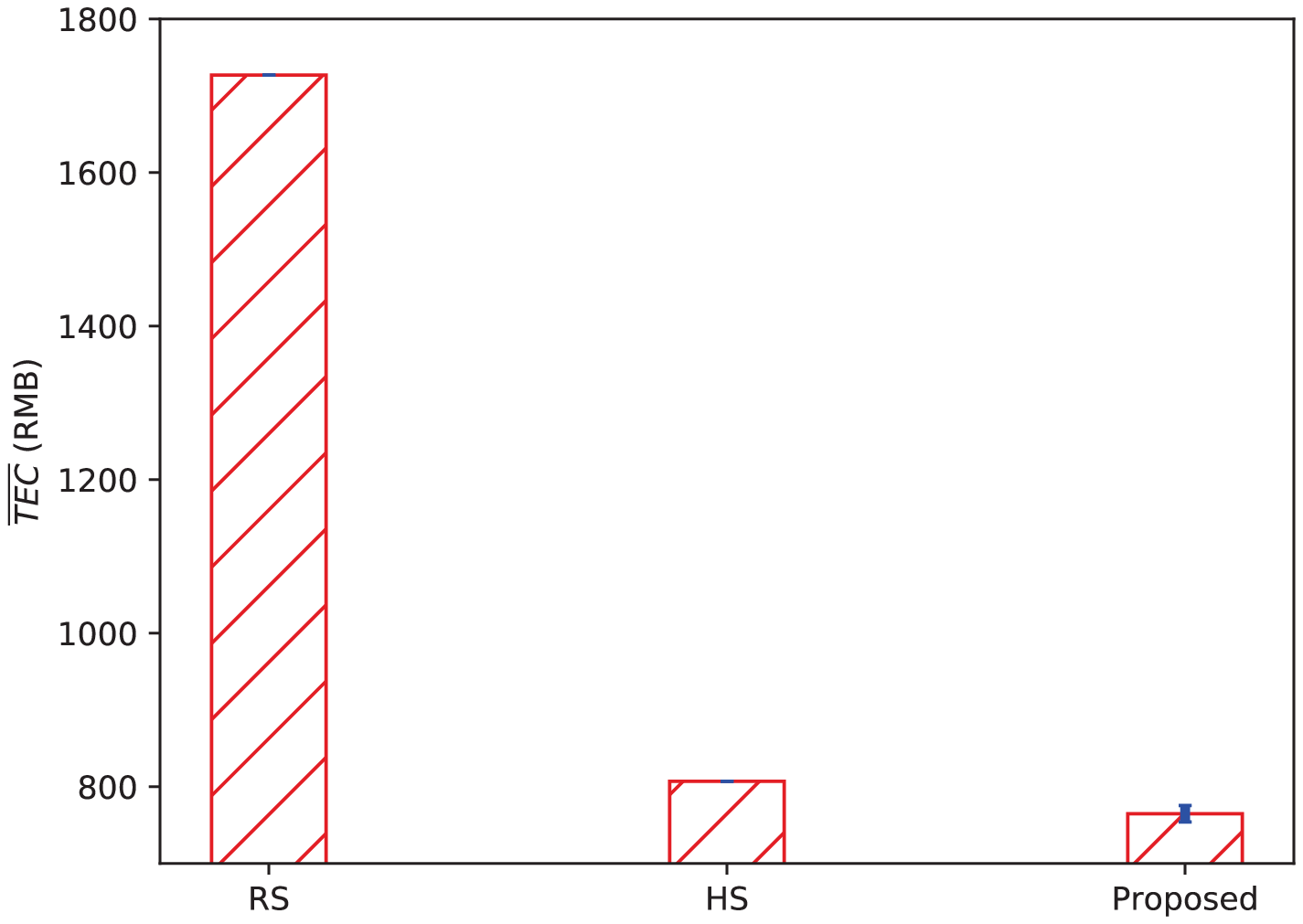}
\end{minipage}
} \\
\subfigure[Total air supply rate]{
\begin{minipage}[b]{0.31\textwidth}
\includegraphics[width=1\textwidth]{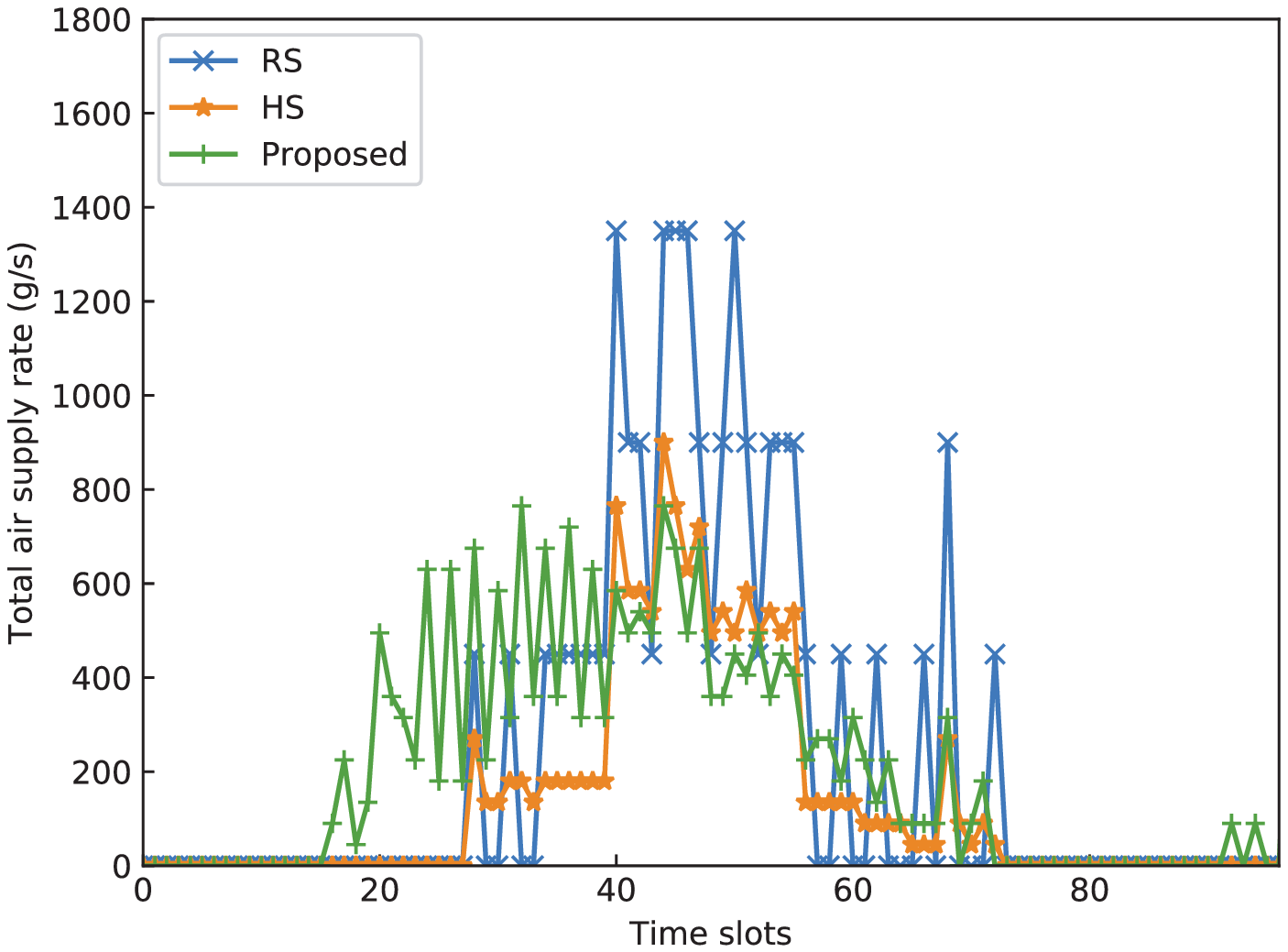}
\end{minipage}
}
\subfigure[Energy cost in each slot]{
\begin{minipage}[b]{0.31\textwidth}
\includegraphics[width=1\textwidth]{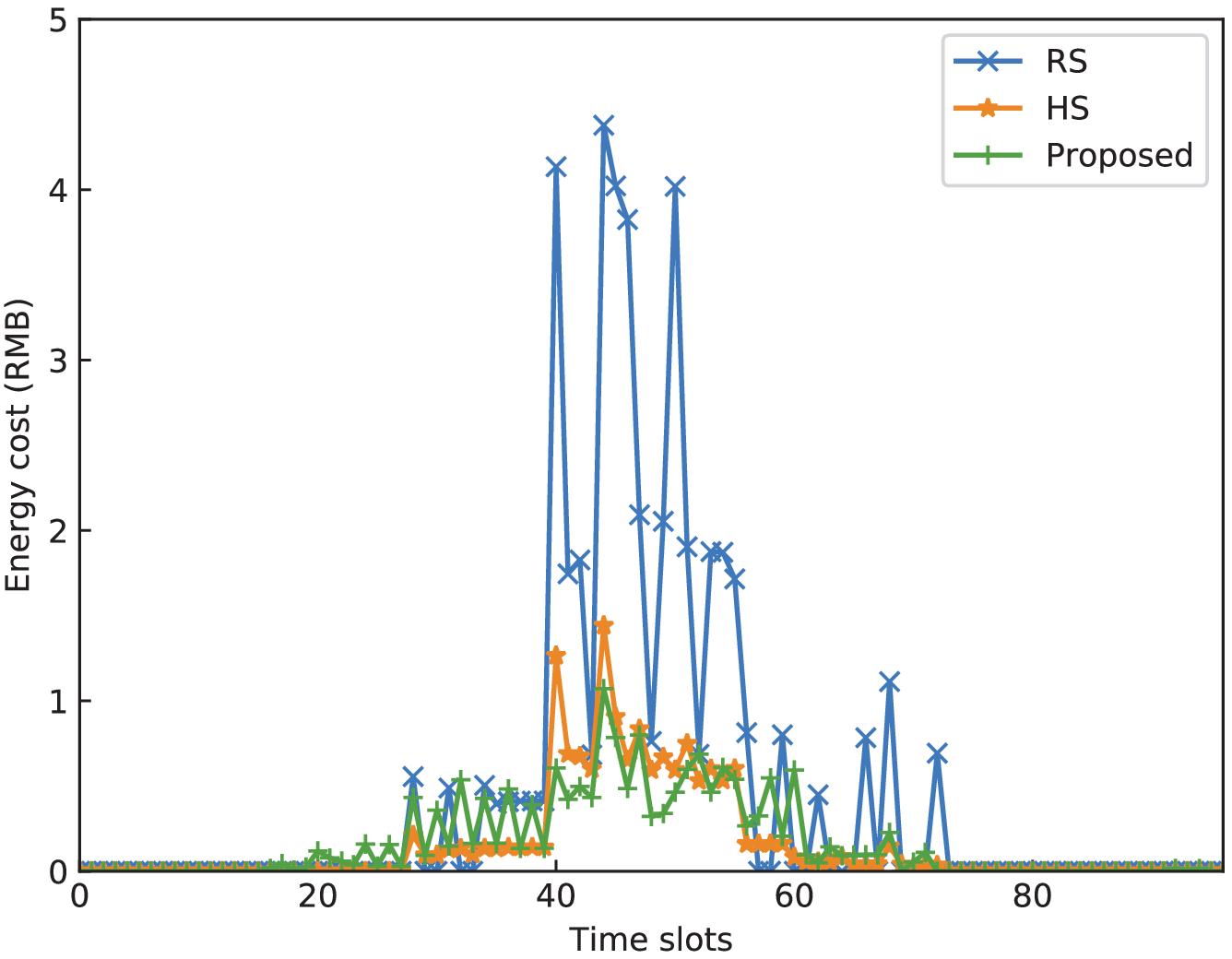}
\end{minipage}
}
\subfigure[Damper position in the AHU]{
\begin{minipage}[b]{0.31\textwidth}
\includegraphics[width=1\textwidth]{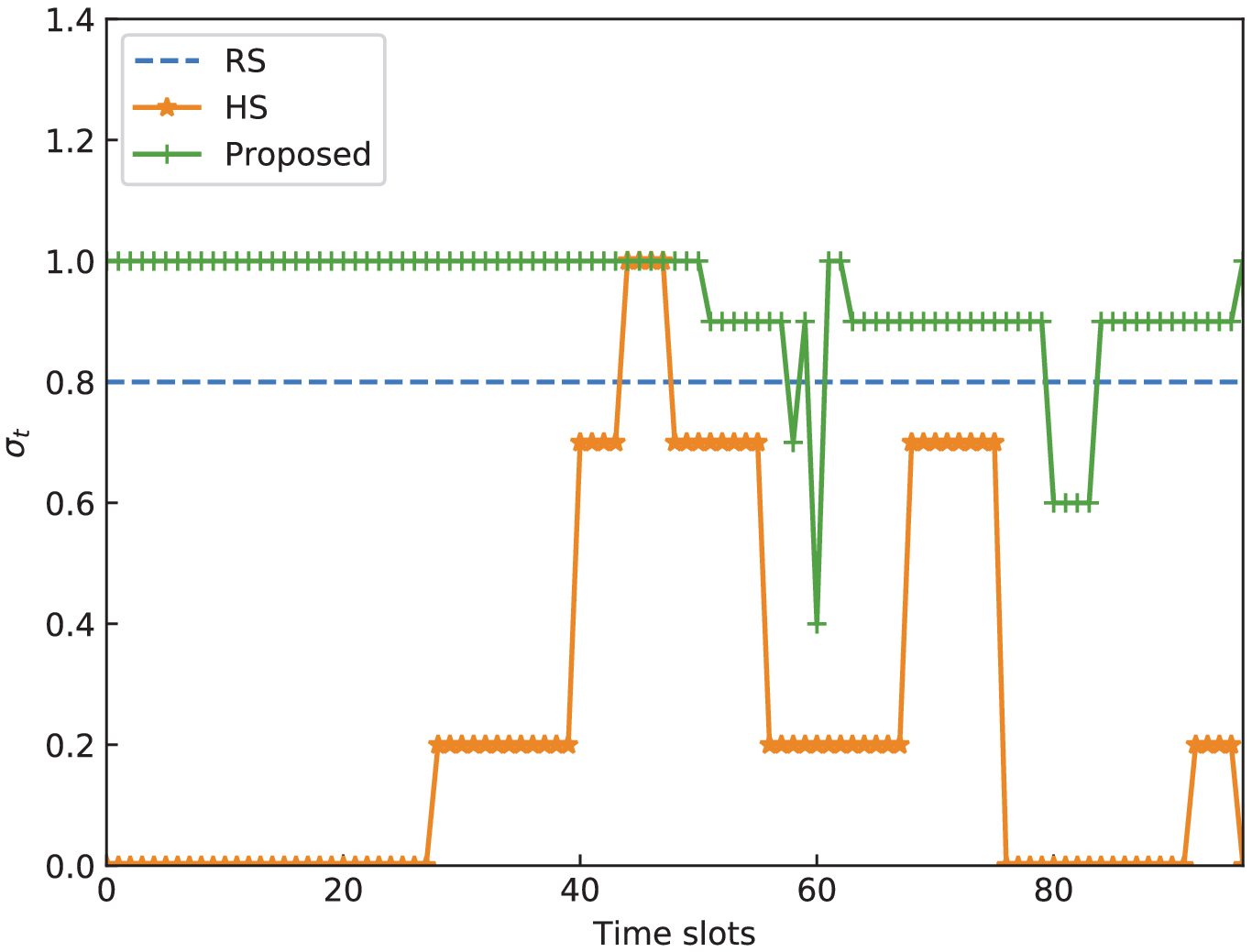}
\end{minipage}
}
\caption{Simulation results of three schemes under the given conditions (i.e., $\overline{ATD}\leq 1.2^oC, \overline{ACD}\leq 40 \text{ppm}$, $\alpha=24$ and $\beta=0.02$)}. \label{fig_6}
\end{figure*}

\subsection{Algorithmic Effectiveness}
We evaluate the effectiveness of the proposed algorithm in this subsection. Since comfortable temperature range and $\text{CO}_2$ concentration range can not always be maintained by $\emph{RS}$ and $\emph{HS}$, performance comparisons are conducted under different conditions as shown in TABLE~\ref{table_2}, where $\overline{ATD}$, $\overline{ACD}$, $\overline{TEC}$ denote the mean value of $ATD$, $ACD$ and total energy cost\footnote{10 experiments (including algorithmic training and testing) are conducted and 95\% confidence interval is considered.}, respectively. It can be seen that the proposed algorithm can satisfy the given conditions and achieve the lowest $\overline{TEC}$ among three schemes. In contrast, the given conditions can not be fully supported by $\emph{RS}$ and $\emph{HS}$ due to the lack of flexibility. To further explain the effectiveness of the proposed algorithm, we provide some simulation results in Fig.~\ref{fig_6}. It can be observed that the proposed algorithm can satisfy the given conditions (i.e., $\overline{ATD}\leq 1.2^oC,~\overline{ACD}\leq 40~\text{ppm}$) and has the lowest $\overline{TEC}$ as shown in Figs.~\ref{fig_6}(a)-(c). The main reason is that the proposed algorithm can implement a flexible coordination among different zones. To be specific, it reduces the total air supply rate when electricity price is high and increases the total air supply rate when electricity price is low, which contributes to the reduction of total energy cost as illustrated in Figs.~\ref{fig_6}(d)-(e). In addition, the proposed algorithm intends to choose a larger $\sigma_t$ while maintaining indoor air quality comfort as shown in Figs.~\ref{fig_6}(b) and (f), which is helpful to save energy.

\subsection{Algorithmic Robustness}


\begin{figure*}
\centering
\subfigure[$\overline{ACD}$]{
\begin{minipage}[b]{0.31\textwidth}
\includegraphics[width=1\textwidth]{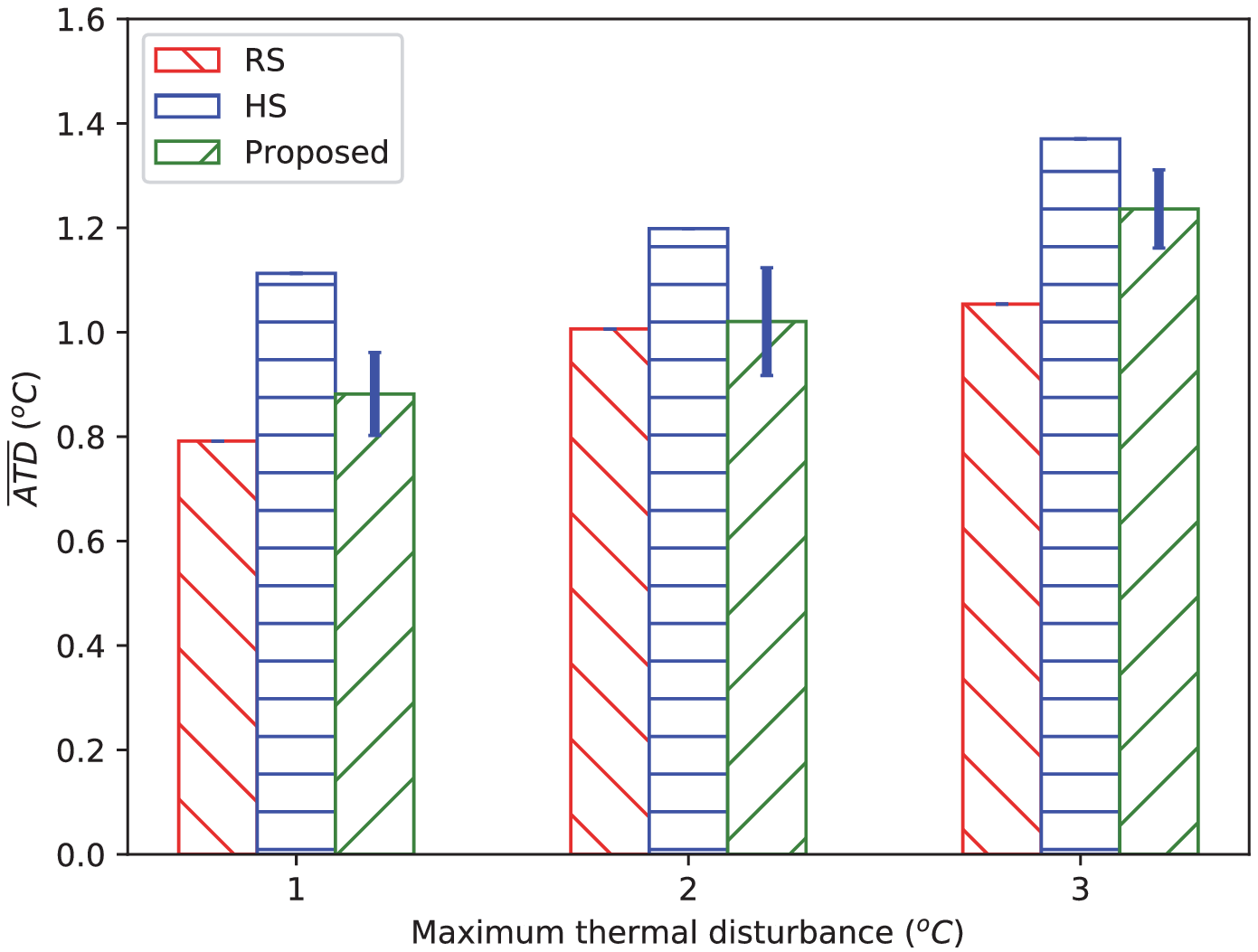}
\end{minipage}
}
\subfigure[$\overline{ACD}$]{
\begin{minipage}[b]{0.31\textwidth}
\includegraphics[width=1\textwidth]{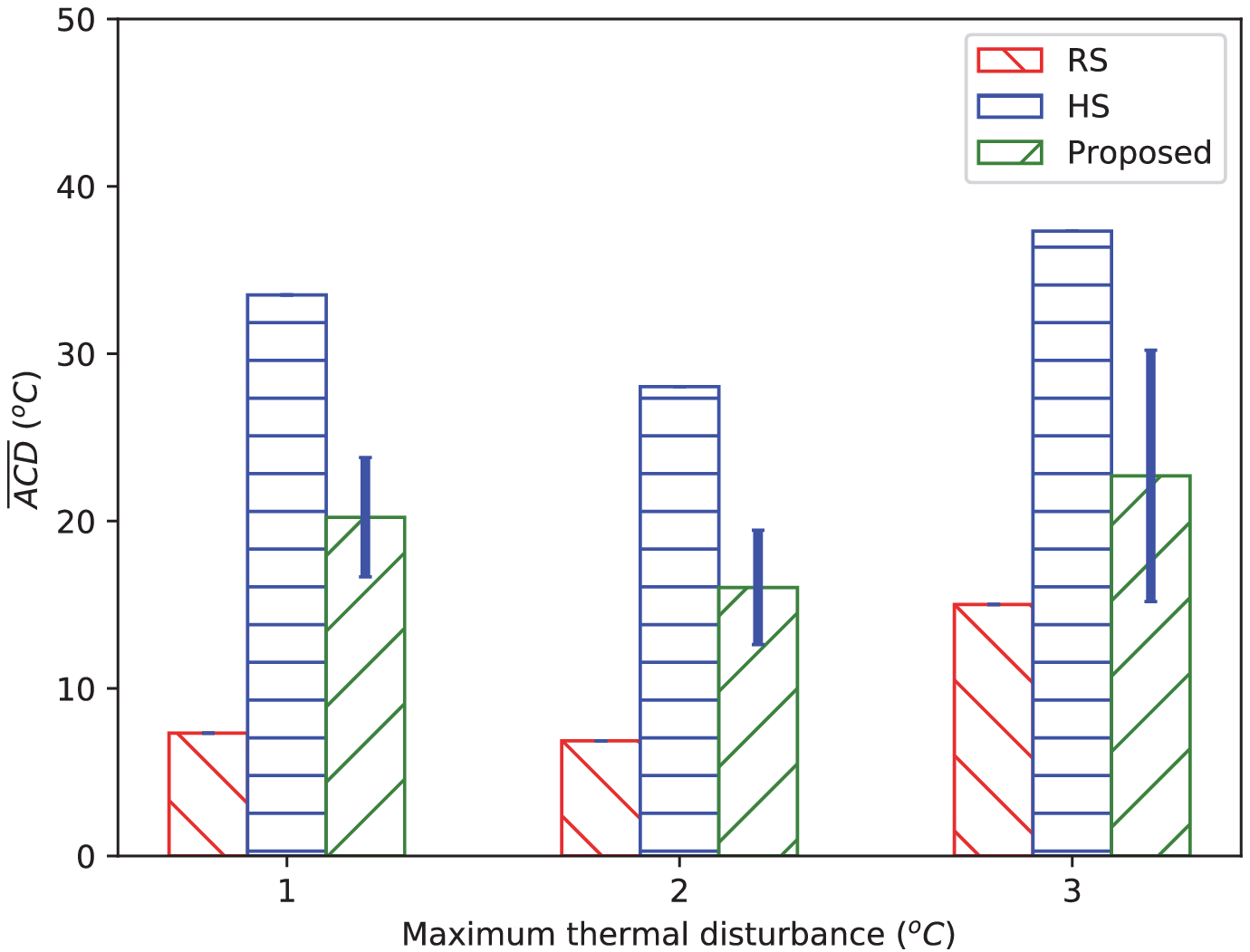}
\end{minipage}
}
\subfigure[$\overline{TEC}$]{
\begin{minipage}[b]{0.31\textwidth}
\includegraphics[width=1\textwidth]{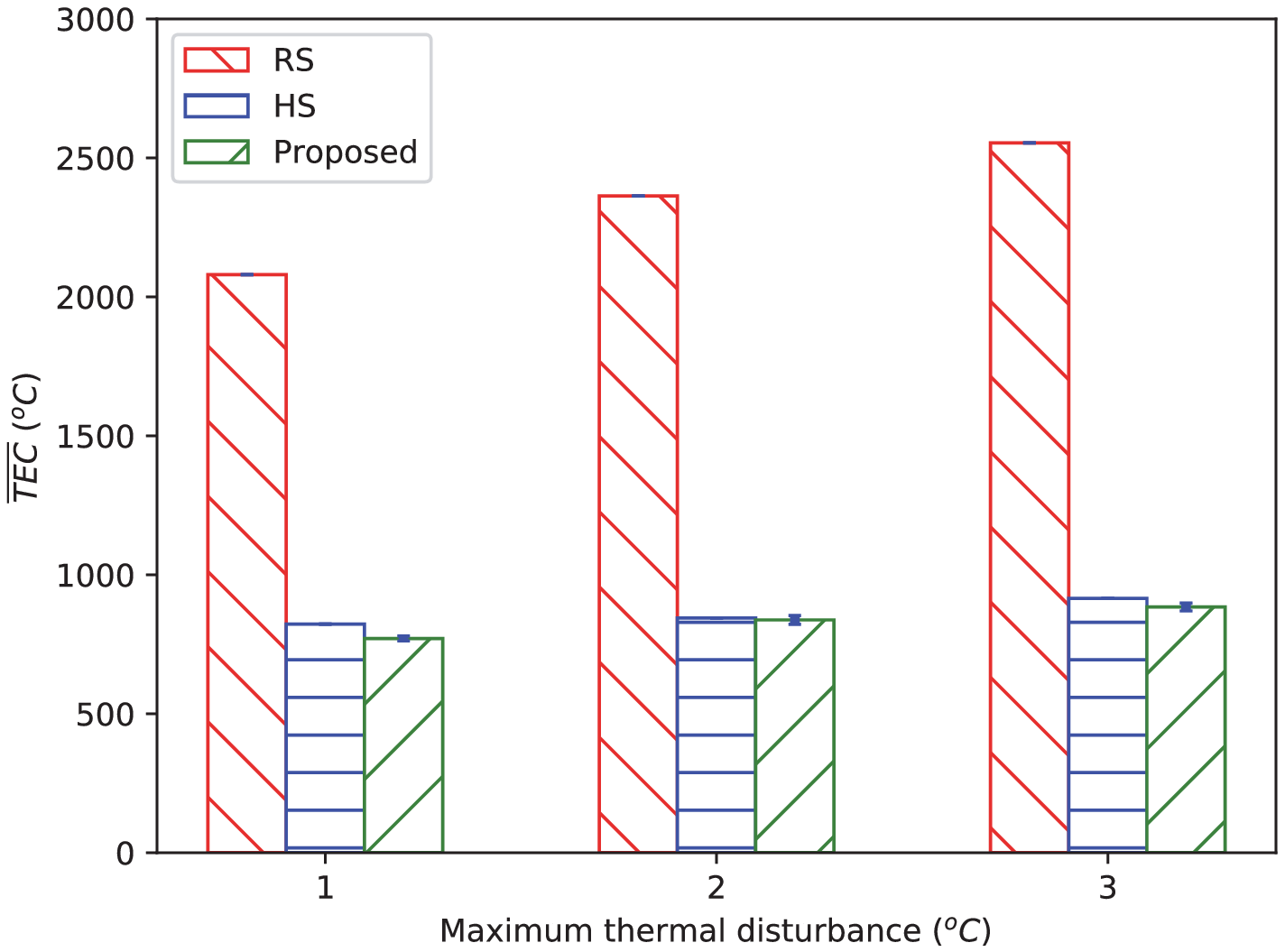}
\end{minipage}
}
\caption{Robustness performance of the proposed algorithm. Note that 95\% confidence interval across 10 runs is considered.} \label{fig_8}
\end{figure*}

\begin{figure*}
\centering
\subfigure[Convergence process]{
\begin{minipage}[b]{0.31\textwidth}
\includegraphics[width=1\textwidth]{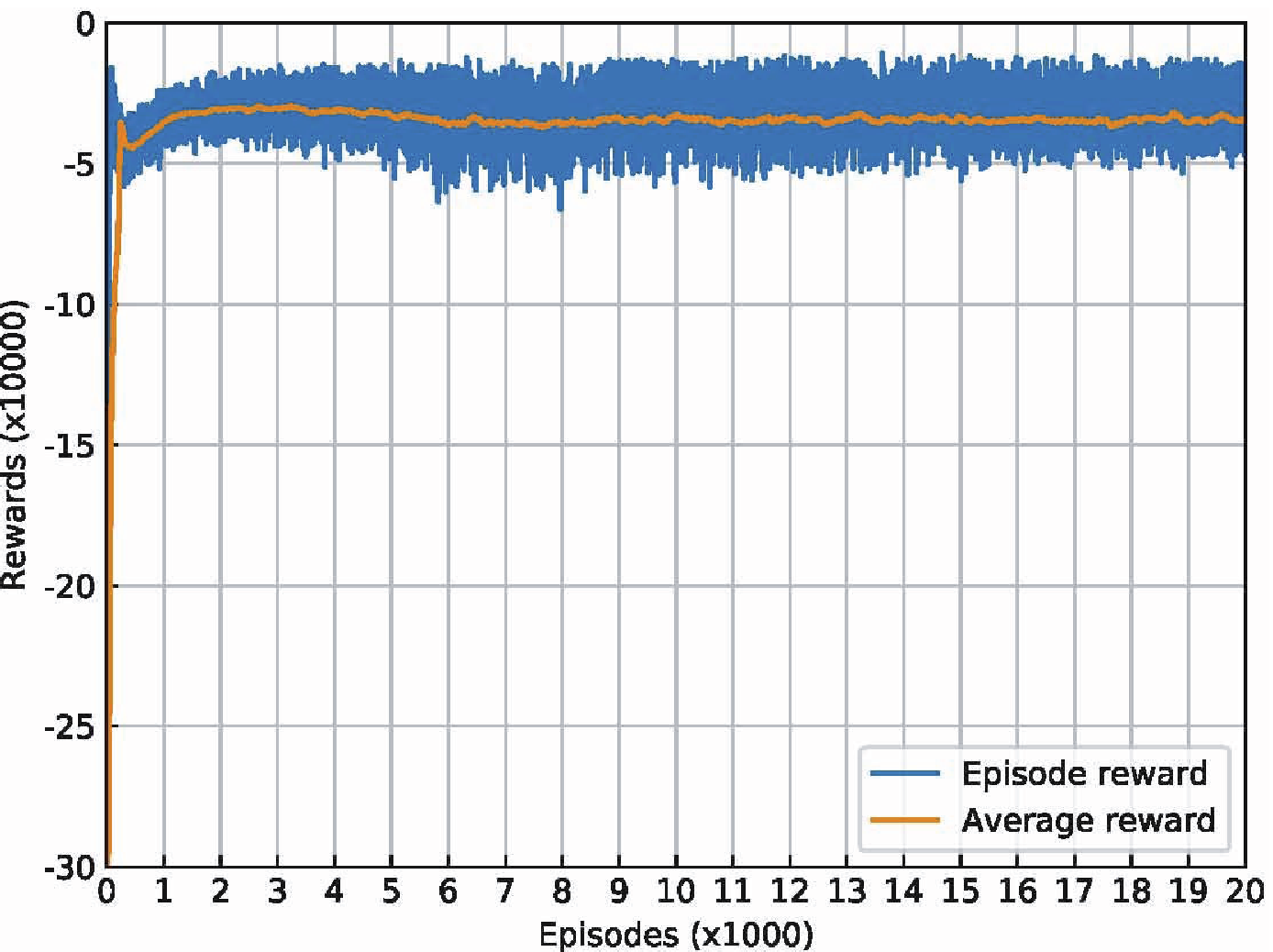}
\end{minipage}
}
\subfigure[$\overline{TEC}$]{
\begin{minipage}[b]{0.31\textwidth}
\includegraphics[width=1\textwidth]{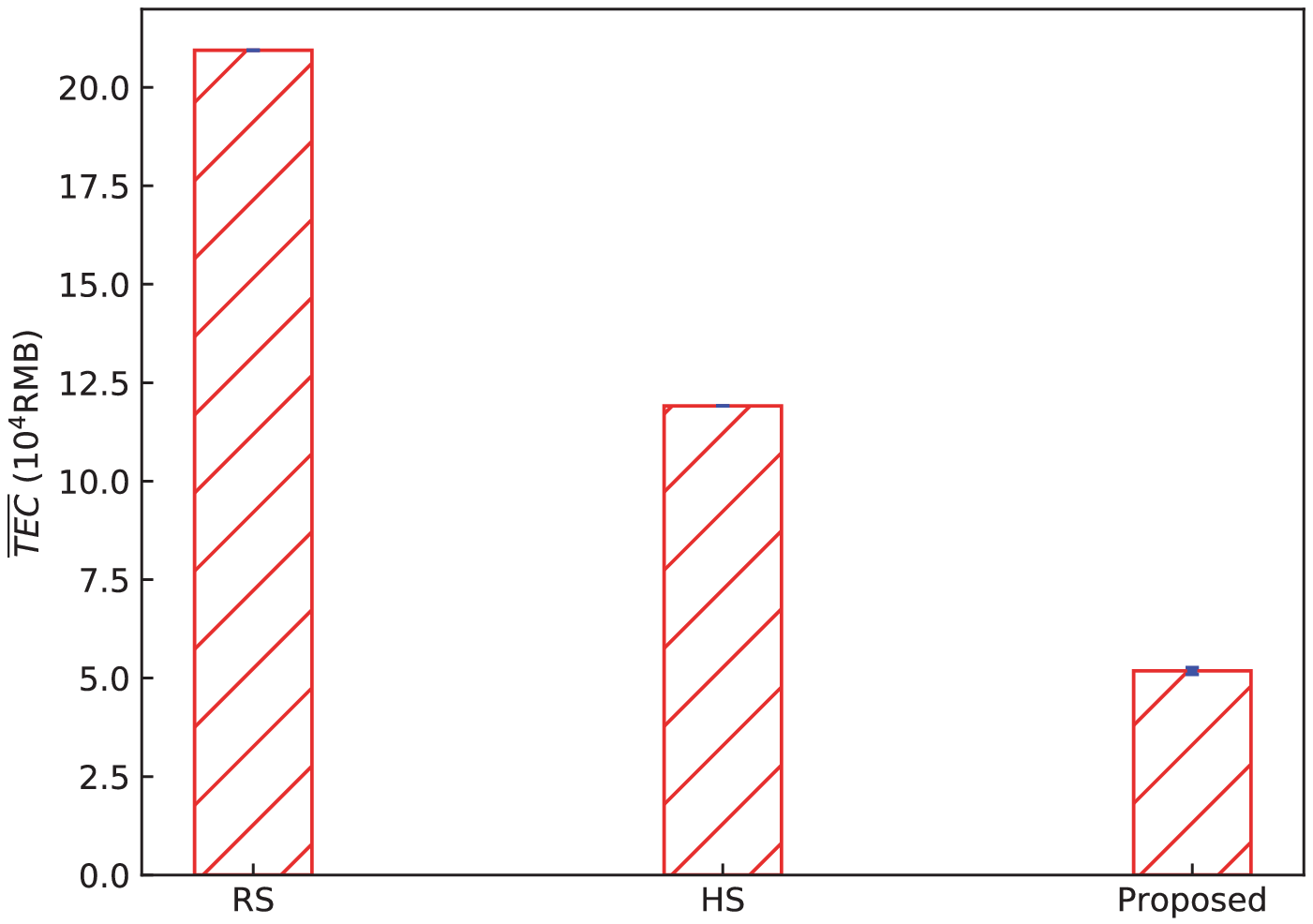}
\end{minipage}
}
\subfigure[Total air supply rate]{
\begin{minipage}[b]{0.31\textwidth}
\includegraphics[width=1\textwidth]{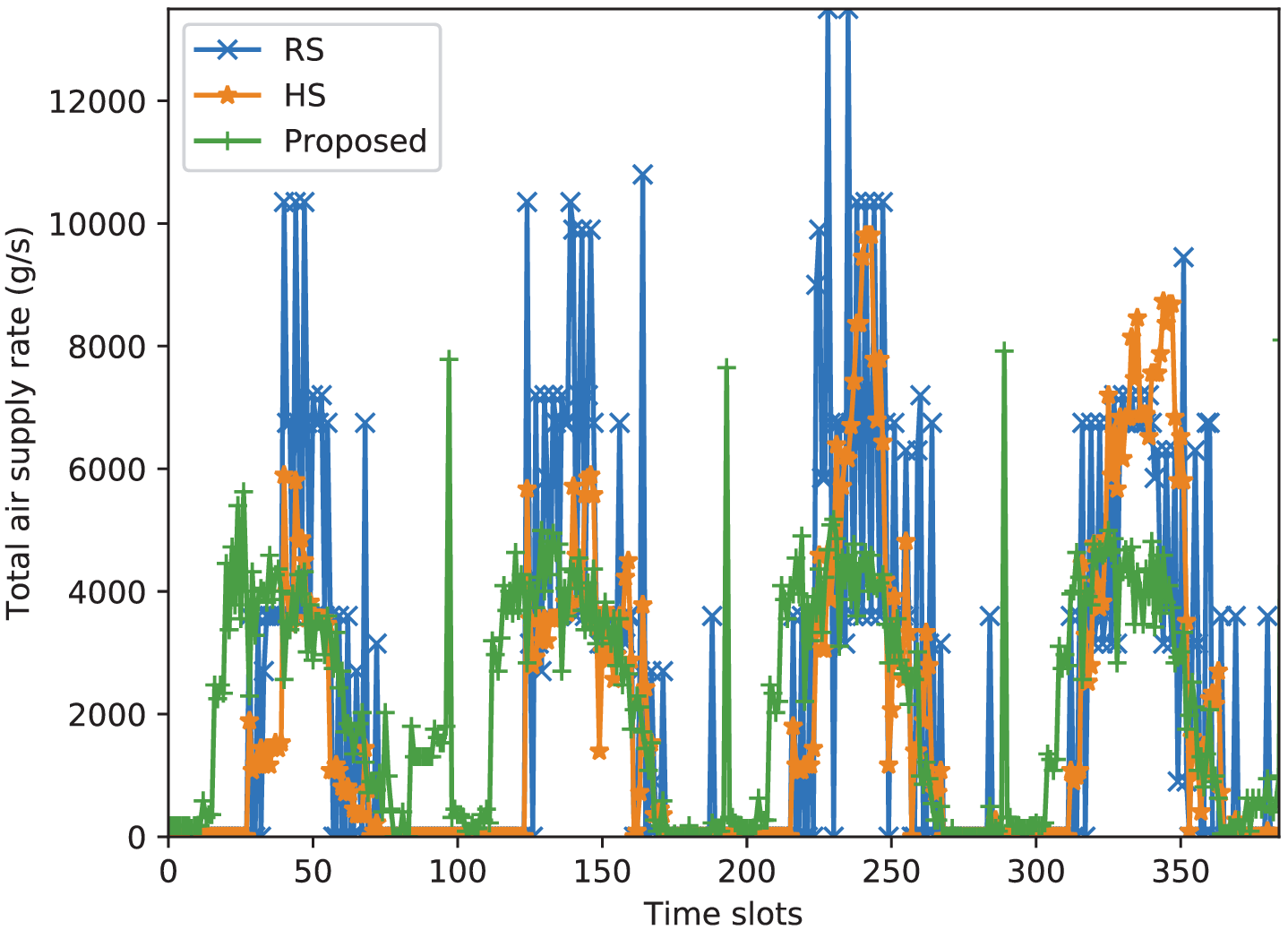}
\end{minipage}
}
\caption{Scalability testing of the proposed algorithm. Note that 30 zones and 95\% confidence interval across 10 runs are considered.} \label{fig_9}
\end{figure*}

To illustrate that the proposed algorithm can be applied to any building thermal dynamics model, we evaluate the robustness of the proposed algorithm when random disturbances are non-zero. Some parameters are configured as follows: $N=4$, $\alpha=24$, $\beta=0.02$, $\zeta=0.9$, $\overline{ATD}\leq 1.4^oC$, $\overline{ACD}\leq 40\text{ppm}$. Moreover, the following building thermal dynamics models are adopted to describe the building environment\cite{LiangTSG2019}, i.e., $T_{i,t+1}=\ell_{i}T_{i,t}+\sum\nolimits_{z\in \mathcal{N}_i} \hbar_{i,z}T_{z,t}+\varpi_i m_{i,t}(T_s-T_{i,t})+\varrho_iT_{t}^{\text{out}}+\Omega_{i,t}$, where $\ell_{i}$,~$\hbar_{i,z}$,~$\varpi_i$,~$\varrho_i$ are RC parameters, $\Omega_{i,t}$ is assumed to follow uniform distribution with parameters $-\upsilon_i$ and $\upsilon_i$. Here, three scenarios (i.e., $\upsilon_i=1^oC,\upsilon_i=2^oC,\upsilon_i=3^oC$) are considered. In Fig.~\ref{fig_8}, it can be seen that the proposed algorithm achieves the best performance under three scenarios, which can justify that the proposed algorithm is robust to random disturbances.

\subsection{Algorithmic Scalability}
As shown in \cite{Iqbal2019}, the MAAC method is scalable to the number of agents as the attention mechanism is adopted. Therefore, the proposed MADRL-based algorithm is also scalable to the number of zones. In this subsection, we intend to verify that the proposed algorithm can still converge and be effective when the number of agents is large. To facilitate the convergence of the proposed algorithm, a larger capacity of experience replay buffer is required for the experiment with a larger number of agents. Due to the lack of enough memory resource, 30 zones are considered for evaluation. Although this number is not large enough, the resulting number of actions at each slot is extremely large, i.e., $11^{31}$. To ensure that all schemes can satisfy the given comfort conditions, i.e., $\overline{ATD}\leq 1.3^oC,~\overline{ACD}\leq 20~\text{ppm}$, hyper-parameters in $\emph{RS}$ and the proposed algorithm are configured as follows, i.e., $\zeta=0.86$, $\alpha=10$, $\beta=0.02$. Fig.~\ref{fig_9}(a) shows the convergence process of the proposed algorithm. It can be observed that the reward received during each episode generally increases. Since soft actor-critic method used in the Algorithm 1 encourages exploration and system parameters are varying in each episode, the episode reward fluctuates within a small range. To show the changing trend of rewards more clearly, we provide the average value of the past 200 episodes, which generally increases and becomes more and more stable. When 20000 episodes are considered, the average training time is about 13 hours. In Fig.~\ref{fig_9}(b), energy cost comparison among three schemes under the given conditions is conducted. It can be seen that the proposed algorithm can reduce $\overline{TEC}$ by 75.25\% and 56.50\% when compared with $\emph{RS}$ and $\emph{HS}$, respectively. The main reason is that the proposed algorithm intends to reduce the total air supply rate of all zones when electricity price is high, which can be depicted by Fig.~\ref{fig_9}(c). Therefore, if an upper limit $\overline{m}<\sum\nolimits_{i=1}^N m_{i}^{M}$ is imposed on the total air supply rate of all zones in practice (i.e., $\sum\nolimits_{i=1}^N m_{i,t}\leq \overline{m}$\cite{Yu2018JIOT}), the above-mentioned performance improvements achieved by the proposed algorithm will decrease. In future work, the impact of $\overline{m}$ on the performance of the proposed algorithm will be investigated.

\section{Conclusion}\label{s5}
In this paper, we proposed a scalable MADRL-based control algorithm to minimize HVAC energy cost in a multi-zone commercial building with the consideration of random zone occupancy, thermal comfort and indoor air quality comfort. The proposed algorithm does not require any prior knowledge of uncertain parameters and can operate in the absence of building thermal dynamics models. The simulation results based on real-world traces showed the effectiveness, robustness, and scalability of the proposed algorithm. In future work, we intend to design DRL-based HVAC control methods for multiple commercial buildings with the consideration of physical constraints related to distribution network\cite{Liu2017} and supply fan\cite{Yu2018JIOT}. In addition, designing energy management methods for commercial building microgrids without knowing building thermal dynamics models is deserved to be investigated.

\end{document}